\documentclass[sigconf]{acmart}

\copyrightyear{2021}
\acmYear{2021}
\setcopyright{acmcopyright}
\acmConference[CCS '21] {Proceedings of the 2021 ACM SIGSAC Conference on Computer and Communications Security}{November 15--19, 2021}{Virtual Event, Republic of Korea.}
\acmBooktitle{Proceedings of the 2021 ACM SIGSAC Conference on Computer and Communications Security (CCS '21), November 15--19, 2021, Virtual Event, Republic of Korea}
\acmPrice{15.00}
\acmISBN{978-1-4503-8454-4/21/11}
\acmDOI{10.1145/3460120.3484742} 

\usepackage{graphicx}
\usepackage{amsmath, amsfonts}
\PassOptionsToPackage{hyphens}{url}\usepackage{hyperref}
\usepackage{xcolor}
\usepackage{url}

\usepackage{csquotes}
\usepackage{subcaption}
\usepackage{diagbox} 
\usepackage[normalem]{ulem}
\useunder{\uline}{\ul}{}
\usepackage[ruled,vlined]{algorithm2e}
\usepackage{multirow}
\usepackage{graphicx}

\newcommand{\rev}[1]{{\color{black} #1}}
\newcommand{\htrev}[1]{{\color{black} #1}}

\newcommand{\para}[1]{{\vspace{1.5pt} \bf \noindent #1 \hspace{4pt}}}   
\newcommand\emily[1]{{\color{black} #1}}
\newcommand\jenna[1]{{\color{black} #1}}

\newcommand\htedit[1]{{\color{black} #1}}

\newcommand{\eg}{{e.g., }}
\newcommand{\ie}{{i.e., }}

\newcommand\adv{$\mathcal{A}$}
\newcommand\sadv{$\mathcal{S}_\mathcal{A}$}
\newcommand\targ{$\mathcal{T}$}
\newcommand\starg{$\mathcal{S}_\mathcal{T}$}

\newenvironment{packed_itemize}{
\begin{list}{\labelitemi}{\leftmargin=1em}
  \setlength{\itemsep}{2pt}
  \setlength{\parskip}{0pt}
  \setlength{\parsep}{0pt}
  \setlength{\headsep}{0pt}
  \setlength{\topskip}{0pt}
  \setlength{\topmargin}{0pt}
  \setlength{\topsep}{0pt}
  \setlength{\partopsep}{0pt}
}{\end{list}}

\begin{document}
\fancyhead{}

\title{``Hello, It's Me'': Deep Learning-based Speech Synthesis Attacks in the Real World}

\author{Emily Wenger}
\authornote{Corresponding author}
\email{ewenger@uchicago.edu}
\affiliation{%
  \institution{University of Chicago}
  \city{}
  \country{}}

\author{Max Bronckers}
\email{mbronckers@uchicago.edu}
\affiliation{%
  \institution{University of Chicago}
  \city{}
  \country{}}

\author{Christian Cianfarani}
\email{crc@uchicago.edu}
\affiliation{%
  \institution{University of Chicago}
  \city{}
  \country{}}

\author{Jenna Cryan}
\email{jennacryan@uchicago.edu}
\affiliation{%
  \institution{University of Chicago}
  \city{}
  \country{}}

\author{Angela Sha}
\email{angelasha@uchicago.edu}
\affiliation{%
  \institution{University of Chicago}
  \city{}
  \country{}}

\author{Haitao Zheng}
\email{htzheng@uchicago.edu}
\affiliation{%
  \institution{University of Chicago}
  \city{}
  \country{}}

\author{Ben Y. Zhao}
\email{ravenben@uchicago.edu}
\affiliation{%
  \institution{University of Chicago}
  \city{}
  \country{}}

\renewcommand{\shortauthors}{Wenger et al.}

\begin{abstract}
  Advances in deep learning have introduced a new wave of voice synthesis
  tools, capable of producing audio that sounds as if spoken by a target speaker.  If
  successful, such tools in the wrong hands will enable a range of powerful
  attacks against both humans and software systems (aka machines). This paper documents
  efforts and findings from a comprehensive experimental study on the impact
  of deep-learning based speech synthesis attacks on both human
  listeners and machines such as
  speaker recognition and voice-signin systems.
  We find that both humans and machines can be reliably fooled by synthetic speech, and that existing
  defenses against synthesized speech fall short. \rev{These 
    findings highlight the need to raise awareness and develop new
    protections against synthetic speech for both humans and machines. }
  
\end{abstract}



\maketitle

\begin{figure*}[t]
  \centering
  \includegraphics[width=0.75\textwidth]{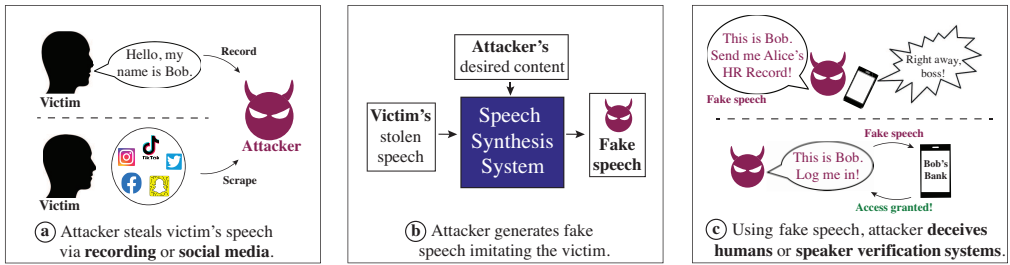}
  \vspace{-0.1in}
  \caption{Workflow of synthesis-based voice spoofing attacks: (a) the attacker obtains voice samples from the
    victim, either by secretly recording them or by downloading
    available media;  (b) the attacker then uses a speech synthesis
    system to generate fake speech, which imitates the victim's voice
    but contains arbitrary, attacker-chosen content; (c) the attacker
uses this fake speech to impersonate the victim, e.g., attempting to access personal or financial information or
    conduct other attacks.}
  \label{fig:attack_pipeline}
\end{figure*}


\vspace{-0.1in}
\section{Introduction}

Our voice conveys so much more than the words we speak. It is a
\rev{fundamental} part of our identity, 
often described as our ``auditory
face''~\cite{voiceper1}.  
Hearing our voice is often enough for a
listener to make inferences about us, such as
gender appearance~\cite{voicegender}, size or strength~\cite{voicesize}, approximate age~\cite{voiceage}, and 
even socioeconomic status~\cite{voiceecon}.

But perhaps the human voice is no longer as unique as we would like to
believe. Recent advances in deep learning have led to a wide range of tools
that produce synthetic speech spoken in a voice of a target speaker, either as
text-to-speech (TTS) tools that transform arbitrary text into spoken words~\cite{wang2017tacotron,jia2018transfer,shen2018natural,
  ping2017deep,hsu2018hierarchical, hu2019neural, arik2018neural, taigman2017voiceloop},
or as voice conversion tools that reshape existing voice samples into the
same content spoken by the target~\cite{rebryk2020convoice,
  wu2020vqvc+,kameoka2018stargan, qian2019autovc,serra2019blow}. In addition to
proprietary systems like Google Duplex, many others are available as
open source software or commercial web services~\cite{lyrebird,resembleai}.


Given the strong ties between our voices and our identities, a tool that
successfully spoofs or mimics our voices can do severe damage in a variety of
settings. First, it could bypass voice-based authentication systems (also
called automatic speaker verification systems) already deployed in automated
customer service phonelines for banks and credit card companies (\eg JP
Morgan Chase and HSBC~\cite{chase_biometrics, hsbc_voice}), as well as user
login services for mobile messaging apps like
WeChat~\cite{wechat_voiceprint}. It would also defeat user-based access
controls in IoT devices such as digital home assistants (\eg Amazon Alexa,
Google Home)~\cite{alexa_voice}. Finally, such tools could directly attack
end users, by augmenting traditional phishing scams with a familiar human
voice. This apparently was the case in a recent scam, where attackers used the mimicked
voice of a corporate CEO to order a subordinate to issue an illegitimate
money transfer~\cite{fakevoice_theft}.



These speech synthesis attacks, particularly those enabled by advances in
deep learning, pose a serious threat to both computer systems and human
beings. Yet, there has been -- until now -- no definitive effort to measure
the severity of this threat in the context of deep learning systems.  Prior
work has established the viability of 
speech synthesis attacks against prior generations of synthesis tools and speaker recognition
systems~\cite{masuko1999security, de2010revisiting,
  kinnunen2012vulnerability, mukhopadhyay2015all}.
Similarly, prior work assessing human vulnerability to speech synthesis
attacks evaluates now-outdated systems in limited
settings~\cite{mukhopadhyay2015all, neupane2019crux}.

We believe there is an urgent need to measure and understand how
deep-learning based speech synthesis attacks impact two distinct entities:
{\em machines} (\eg automated software systems) and {\em humans}. Can such
attacks overcome currently deployed speaker recognition systems in
security-critical settings? Or can they compromise mobile systems such as
voice-signin on mobile apps? Against human targets, can synthesized speech samples
mimicking a particular human voice successfully convince us of their authenticity?


In this paper, we describe results of an in-depth analysis of the threat
posed to both machines and humans by deep-learning speech synthesis
attacks. We begin by assessing the susceptibility of modern speaker
verification systems (including commercial systems Microsoft Azure, WeChat,
and Alexa) and evaluate a variety of factors affecting attack success. To
assess human vulnerability to synthetic speech, we perform multiple user
studies in both a survey setting and a trusted context. Finally, we assess
the viability of existing defenses in defending against speech synthesis
attacks. All of our experiments use publicly available deep-learning speech
synthesis systems, and our results highlight the need for new defenses
against deep learning-based speech synthesis attacks, \rev{for
      both humans and machines.}



\para{Key Findings.} Our study produces several key findings:
  \begin{packed_itemize}\vspace{-0.05in}
  \item Using a set of comprehensive experiments over $90$ different speakers, we
    evaluate and show that DNN-based speech synthesis tools are highly
    effective at misleading modern speaker recognition systems ($50-100\%$ success).
  \item Our experiments find that given a handful of attempts, synthesized
    speech can mimic $60\%$ of speakers in real world speaker recognition
    systems: Microsoft Azure, WeChat, and Amazon Alexa. 
  \item A user survey of $200$ participants shows humans can distinguish
    synthetic speech from the real speaker with $\sim$50\% accuracy for unfamiliar voices
    but near 80\% for familiar voices.
  \item An interview-based deception study of $14$ participants shows that,
    in a more trusted setting, inserted synthetic speech successfully deceives
    the large majority of participants.
  \item Detailed evaluation of $2$ state-of-the-art defenses shows that they fall
    short in their goals of either preventing speech synthesis or reliably
    detecting it, highlighting the need for new defenses. 
    \vspace{-0.in}
  \end{packed_itemize}
It is important to note that speech synthesis is intrinsically about producing
audible speech that sounds like the target speaker to humans and machines
alike. This is fundamentally different from adversarial attacks that perturb
speech to cause misclassification in speaker recognition
systems~\cite{chen2019real,li2020practical, kreuk2018fooling}. Such attacks do not affect human listeners, and could be
addressed by developing new defenses against adversarial examples.

\vspace{-0.05in}
\section{Background}
\label{sec:back}

In this section, we first describe current trends in speaker 
recognition technology and voice synthesis
systems, followed by voice-based spoof attacks. Finally, we briefly
summarize defenses proposed to combat synthetic
speech. 

\vspace{-0.08in}
\subsection{Voice-Based User Identification}
\vspace{-0.05in}
\para{How Humans Identify Speakers via Voice.} The unique
characteristics of each person's vocal tract create their distinct
voice. Humans use these vocal characteristics to identify people by
voice~\cite{lopez2013vocal}. Though human speaker
identification is imperfect, it is highly accurate and has inspired
the construction of speaker recognition systems for security purposes~\cite{sharma2019talker}.

\para{Automated User Verification by Machines.} Recently, speaker
recognition has become a popular alternative to other biometric authentication methods~\cite{rosenberg1976automatic}. Speaker recognition
systems capture characteristics of a speaker's voice and compare them
to enrolled speaker profiles. If there is a match, the recognition
system grants the speaker access. Early speaker recognition systems
(1970s-2010s) used parametric methods like Gaussian Mixture Models,
while more recent systems (2014 onward) use deep learning models, which reduce overhead and improve accuracy~\cite{furui1981cepstral, reynolds2000speaker,
  variani2014deep, snyder2016deep}.

Speaker recognition is used in numerous settings, from bank
customer identification to mobile app login and beyond~\cite{chase_biometrics, hsbc_voice,
  wechat_voiceprint}. Recently, virtual assistants like Alexa and
Google Assistant have begun to use speaker recognition to customize system behavior~\cite{alexa_voice,
  google_home}.   Speaker recognition systems are either {\em
  text-dependent} or {\em text-independent}~\cite{bimbot2004tutorial,
  heigold2016end}. Text-dependent systems use the same, speaker-specific
authentication phrase for both enrollment and login. Text-independent
systems are content-agnostic.

\vspace{-0.1in}
\subsection{Speech Synthesis Systems}


Synthetic speech is produced by a non-human source
(i.e. a computer) and imitates the {\em sound} of a human voice. Efforts to produce
electronic synthetic speech go back to 1930s, where Homer Dudley developed
the first vocoder~\cite{first_speech}. Since then, systems like
Festvox~\cite{festvox} have used Gaussian Mixture Models (GMM) to improve the
quality -- but not the speed -- of speech synthesis. The recent deep learning
revolution has catalyzed growth in this field.


\para{DNN-Based Speech Synthesis.} Numerous deep neural network (DNN) based speech synthesis systems have been proposed~\cite{wang2017tacotron, shen2018natural, jia2018transfer,hsu2018hierarchical,
  hu2019neural, taigman2017voiceloop,serra2019blow,kameoka2018stargan,qian2019autovc,
  ping2017deep, rebryk2020convoice, arik2018neural, wu2020vqvc+, qian2020unsupervised}. They
can be divided into two categories: text-to-speech
(TTS) and voice conversion (VC).

TTS systems transform arbitrary text into words spoken in the voice of a target
speaker~\cite{wang2017tacotron,jia2018transfer,shen2018natural,
  ping2017deep,hsu2018hierarchical, hu2019neural, arik2018neural,
  taigman2017voiceloop}. In contrast, VC 
systems take two voice samples -- an attacker and target -- and
output a speech sample in which {\em content} from the attacker is
spoken in the {\em voice} of the target~\cite{rebryk2020convoice,
  wu2020vqvc+,kameoka2018stargan, qian2019autovc,serra2019blow, qian2020unsupervised}. Both
TTS and VC produce the same output: a synthetic version of the target's
voice, speaking words chosen by the attacker.

\para{Efficacy and Availability.} Many DNN-based speech synthesis
systems report impressive speech ``realism'' metrics, indicating
significant improvement over classical systems. Supporting
evidence of DNN synthesis performance comes from
real-world anecdotes. DNN-based synthetic speech has been successfully
used at least one in highly profitable attack~\cite{fakevoice_theft}. Google's new
scheduling assistant voice is so realistic that Google was instructed
to announce when it was being used on phone calls~\cite{google_duplex}.

Some DNN synthesis systems (and their training datasets) remain internal to
companies, but many systems are available on Github~\cite{github_sv2tts, github_autovc, github_mozilla_tts,
  github_tensorflow_tts}. For the less tech-savvy, online
services will perform voice cloning for a fee~\cite{lyrebird,
  resembleai}. This combination of speech synthesis efficacy and
availability is both exciting and worrisome. 

\para{Misuse of Speech Synthesis.} There are many positive uses for speech
synthesis technology, such as giving voices to the mute, aiding spoken
language translation, and increasing human trust of 
helper robots~\cite{anumanchipalli2019speech, krishna2020speech,
  di2019adapting, kuo2009age,tamagawa2011effects}. However, our work focuses
on the ``shadow side'' of these uses -- generating synthetic speech with malintent
to deceive both humans and machines.

\vspace{-0.1in}
\subsection{Voice-based Spoofing Attacks}

In this work, we focus specifically on {\em spoofing attacks} against
voice-based user identification, in which an attacker mimics a
target's voice to steal their identity. \emily{A parallel line of work
explores {\em adversarial attacks}, in which an adversary adds inaudible perturbations to speech to fool speaker recognition systems~\cite{chen2019real,
  li2020practical, kreuk2018fooling}. While powerful, adversarial
attacks differ from spoofing attacks because they do not mimic
the target and so pose no threat to humans.}

Figure~\ref{fig:attack_pipeline} gives a high-level overview of
spoofing attacks. There are several techniques the adversary could use, and these are taxonomized in
Table~\ref{tab:attacks}.  Prior work has found that all spoofing
techniques -- replay, impersonation, and synthesis -- can reliably
fool machine-based voice recognition systems, \emily{but only a few works
  have investigated the threat posed to humans}. Here, we summarize
prior work that studied these spoofing attacks.

\begin{table}
\centering
\resizebox{\linewidth}{!}{%
\begin{tabular}{c|c|c|c|c} 
\hline
\multirow{2}{*}{\textbf{Attack Type}} & \multirow{2}{*}{\textbf{Description}} & \multirow{2}{*}{\begin{tabular}[c]{@{}c@{}}\textbf{Arbitrary}\\\textbf{Content}\end{tabular}} & \multicolumn{2}{c}{\textbf{Measurement Studies }} \\ 
\cline{4-5}
 &  &  & \textit{vs. Machines} & \textit{vs. Humans} \\ 
\hline
Replay & \begin{tabular}[c]{@{}c@{}}\textit{Play pre-recorded}\\speech from victim.\end{tabular} & No & \cite{alegre2014re,janicki2016assessment,wu2015asvspoof, kinnunen2017asvspoof} & - \\ 
\hline
\begin{tabular}[c]{@{}c@{}}Human\\Impersonation\end{tabular}
                                      & \begin{tabular}[c]{@{}c@{}}\textit{Human
                                          actor} \\imitating
                                          victim.\end{tabular} &
                                                                 Limited & \cite{hautamaki2017acoustical,gao2021detection,vestman2020voice, shirvanian2014wiretapping,lau2004vulnerability} & \cite{hautamaki2017acoustical} \\ 
\hline
\begin{tabular}[c]{@{}c@{}}Synthesis\\(Classical)\end{tabular}
                                      & \begin{tabular}[c]{@{}c@{}}\textit{Clone}
                                          victim's speech \\(GMM-based)\end{tabular} & Yes & \cite{masuko1999security,de2010revisiting,kinnunen2012vulnerability,breakability,mukhopadhyay2015all} & \cite{mukhopadhyay2015all} \\ 
\hline
\begin{tabular}[c]{@{}c@{}}Synthesis\\(DNN)\end{tabular}
                                      & \begin{tabular}[c]{@{}c@{}}\textit{Clone}
                                          victim's speech
                                          \\(DNN-based)\end{tabular} &
                                                                       Yes
                                                                              &
                                                                                \cite{partila2020deep}, 
                                                                                {\bf
                                                                                this
                                                                                work}&{\bf this work}
  \\ 
\hline
\end{tabular}
}
\caption{Taxonomy of spoofing attacks against voice-based
  authentication and measurement studies on these attacks.}
\label{tab:attacks}\vspace{-0.25in}
\end{table}


\para{Spoofing Attacks Against Machines.} We first summarize prior
work measuring machines' vulnerability to spoofing attacks. 
\begin{packed_itemize}
  \item {\bf Record-and-Replay:} In a replay attack, an adversary records a
victim's exact speech and replays it to fool a target speaker
recognition system~\cite{alegre2014re, janicki2016assessment}. The ASVspoof Challenge~\cite{wu2015asvspoof,
  kinnunen2017asvspoof} has investigated this attack
extensively. Replay attacks have high overhead since
the attacker must obtain specific recordings of the victim. Furthermore,
this attack is constrained by the content of the victim's available
recordings.

\item {\bf Human Impersonation:} Human voice actors can impersonate
others' voices to great success, and well-crafted impersonation spoofing attacks reliably fool speaker recognition
systems~\cite{hautamaki2017acoustical,gao2021detection,vestman2020voice,shirvanian2014wiretapping,lau2004vulnerability}. These
attacks have even defeated HSBC's speaker recognition-based security~\cite{twin_bank_attack}. While effective,
these attacks have high overhead and limited versatility due to their
dependence on human talent. 

\item {\bf Machine Synthesis (Classical):} Most prior work uses GMM-based speech synthesis systems (e.g., 
Festvox~\cite{festvox}) to attack public, GMM-based speaker
recognition systems~\cite{masuko1999security, de2010revisiting,
  kinnunen2012vulnerability, mukhopadhyay2015all}. \htedit{A recent work
takes a ``real-world'' focus by testing a small set of
synthetic speech generated by Festvox against five mobile apps that
support voice authentication, and reports 96\%+
success~\cite{breakability}. 
However, the 
efficacy of classical synthesis attacks against modern speaker 
recognition systems remains unclear. }

\item \para{Machine Synthesis (DNN-based):}
\htedit{To the best of our knowledge, only one
work~\cite{partila2020deep} has examined the performance of
DNN-based synthesis attacks. It performed preliminary tests by running
10 synthesized samples for 6 speakers (generated
by~\cite{jia2018transfer}) against three locally-trained speaker recognition
prototypes. It produced vague conclusions: these speaker recognition
prototypes produce more errors when running on synthesized
speech compared to clean (non-synthesized) speech. }



\end{packed_itemize}

\emily{\para{Spoofing Attacks Against Humans.} Existing work assessing human
susceptibility to spoofing only evaluates impersonation and classical
synthesis attacks. The single impersonation attack paper found
that humans can be fooled by actors pretending to be older or younger than they really
are~\cite{hautamaki2017acoustical}.} The first classical synthesis
attack measurement paper~\cite{mukhopadhyay2015all} uses a
traditional survey format and finds that users correctly distinguish
between real and Festvox-synthesized voices (imitating the real speaker) about $50\%$ of the
time, regardless of their familiarity with the real
speaker. A follow-up study to this~\cite{neupane2019crux} uses the
same data and survey format but includes fNIRS brain scanning technology to measure
participants' neural activity. They find no statistically significant
differences in neural activity when real or synthetic
speakers are played. 


\vspace{-0.06in}
\subsection{Defending Against Synthesized Speech}

Numerous defenses have been proposed to defend speech recognition
systems against synthetic speech attacks. While most have focused on
detecting synthetic speech or speakers~\cite{zhang2017hearing,
  wang2019voicepop, chen2017you, catcher, albadawy2019detecting,
  wang2020deepsonar, ahmed2020void, gao2010audio, shirvanian2020voicefox}, recent work has pointed
towards a new defense direction: preventing unauthorized speech
synthesis~\cite{huang2020defending}. \rev{We discuss and evaluate
  representative defenses in \S\ref{sec:existingdef}}.

\vspace{-0.07in}
\section{Methodology}
\label{sec:threat}


No comprehensive study exists today that studies the threat
posed by DNN-based speech synthesis to \htedit{software-based 
speaker recognition systems and human users}. Our work addresses this
critical need, and outlines future work needed to mitigate the resulting
threat. Here, we describe the threat model, and the methodology, tools and
datasets used by our analysis. 

\subsection{Threat Model and Assumptions}


In DNN-based speech synthesis attacks, the
adversary \adv{}'s goal is to steal a target \targ{}'s identity by
imitating their voice. To do so, \adv{} first collects a set of 
speech samples \starg{} from \targ, either by secretly recording their
speech in a public setting or by extracting the audio from public
video/audio 
clips.  When \adv{} knows \targ{} personally, these speech clips could
also be
obtained from private media. Next, \adv{} inputs \starg{}
to a speech synthesis system, which produces synthesized or fake voice
samples \sadv. In this case, \sadv{} should sound like \targ{} but
contain arbitrary speech content chosen by \adv{}. 



We make the following assumptions about the adversary \adv{}:
\begin{packed_itemize}\vspace{-0.02in}
\item \adv{} only needs a small volume of speech samples from \targ, i.e., less than $5$
  minutes of speech in total; 
\item  \adv{} \htedit{directly} uses a publicly available, DNN-based voice synthesis
  system to generate synthetic speech \sadv{}; 
  \item \adv{} seeks to generate fake voice samples \sadv{} that make either {\em humans} or {\em
  machines} believe that they are interacting with \targ{}. \vspace{-0.02in}
\end{packed_itemize}

\subsection{Overview of Experiments}
We conduct a measurement study to explore the real-world threat posed to
both {\bf machines} and {\bf humans} by today's publicly available,
DNN-based speech synthesis systems.   These include: 

\begin{packed_itemize}\vspace{-0.02in}
\item Empirical experiments to examine whether synthetic speech can
  fool speaker
  recognition (SR) systems, aka machines (\S\ref{sec:machines});
 \item User studies to explore human
   susceptibility to synthetic speech under
  multiple interactive scenarios (\S\ref{sec:humans});
 \item Empirical experiments to assess the efficacy of existing defenses against DNN-based
synthesis attacks (\S\ref{sec:existingdef}). \vspace{-0.02in}
\end{packed_itemize}
In the following, we describe the DNN synthesis and SR systems as well as speaker
datasets used by our experiments.

\subsection{DNN-based Synthesis Systems Studied} 
We consider "zero shot" systems (i.e., those requiring < 5 minutes of
target data to run synthesis)  and focus on peer-reviewed, published papers with public code
implementations and pre-trained models. We tested numerous synthesis systems including~\cite{serra2019blow, taigman2017voiceloop, arik2018neural,
  kameoka2018stargan, qian2019autovc, jia2018transfer}, but found that many did not generalize well to unseen
speakers (i.e., those not in the training dataset). Generalization is
critical to low-resource attackers like the ones defined by our threat model,
as it allows flexibility in target choice. In the end, we
chose two systems that performed best on unseen speakers: SV2TTS ~\cite{jia2018transfer}, a TTS
system based on Google's Tacotron, and AutoVC~\cite{qian2019autovc},
an autoencoder-based voice conversion system. 

\para{SV2TTS.} This is a zero-shot, text-independent voice
conversion system requiring only five seconds of target
speech~\cite{jia2018transfer}. It combines
three earlier works from Google: an LSTM speaker encoder
~\cite{wan2018generalized}, the DNN speech synthesis network Tacotron
2 and the WaveNet vocoder. 
\htedit{Instead of training the model ourselves}, we use a well-known
public implementation~\cite{corentin2019thesis}.  In this case, the encoder is pre-trained
on the  VoxCeleb1/2~\cite{nagrani2020voxceleb} and
LibriSpeech \texttt{train}~\cite{libri}  datasets,  and the synthesis network is pre-trained on
LibriSpeech \texttt{train}, both following the original paper's
setting~\cite{jia2018transfer}.
%

\para{AutoVC.}  The second system is a zero-shot style transfer
autoencoder network that performs text-independent voice conversion~\cite{qian2019autovc}.
Its encoder bottleneck disentangles speaker characteristics from
speech content to facilitate speech synthesis. Like SV2TTS, it relies on
~\cite{wan2018generalized} and WaveNet for its speaker encoder and speech
vocoder, respectively.  \htedit{We also use the publicly available
implementation provided by}~\cite{github_autovc}, where the speaker encoder is
pre-trained on VoxCeleb1 and
LibriSpeech \texttt{train}, and the autoencoder is pre-trained on VCTK, again
following~\cite{qian2019autovc}'s original setting.

\vspace{-0.05in}
\subsection{Speaker Recognition (SR) Systems Studied}
\label{subsec:sr} 
To explore the real-world threat of speech synthesis
attacks on machines,  we choose  {\em four} state-of-the-art SR 
systems. These include both publicly available and
proprietary systems.


\para{Resemblyzer.}
Resemblyzer~\cite{github_resemblyzer} is an
open source DNN speaker encoder \htedit{widely used by recent literature}. It is 
 trained on VoxCeleb 1/2 and LibriSpeech \texttt{train} using
 the generalized end-to-end loss~\cite{wan2018generalized}. 
 Each speaker is enrolled into the system database using about 30 seconds
 of their speech, creating an embedding that represents their
 identity. To recognize an incoming speaker, the system computes the embedding and compares it to embeddings in the
 database using cosine similarity. 



\para{Microsoft Azure.} Microsoft Azure's cloud platform includes a text-independent speaker recognition API~\cite{azure_speaker_rec}. Speakers are
enrolled using 20 seconds of speech data, and speaker verification
queries are made via the API.  This system is certified by several 
international bodies, e.g., the Payment Card Industry (PCI), HIPAA, and International Standards Organization (ISO).

\para{WeChat.} WeChat is a popular mobile messaging and payment
platform 
that offers a text-dependent "voiceprint" login for authentication. Users create their voiceprint by repeating an eight digit
number provided by the app. This same number is used for each
subsequent voiceprint login.  Users have a cap of six voice login
attempts per day before the app enforces password authentication~\cite{wechat_doc}.

\para{Amazon Alexa.} Alexa is a popular virtual assistant embedded in Amazon's smart speakers. 
Alexa uses ``voice profiles'' to customize user
interactions 
and control access to sensitive apps like
email and calendar~\cite{alexa_voice}. Voice profiles control
access to sensitive information in Alexa 3rd party apps, such as
payment (Uber) and phone account management (Vodaphone)~\cite{alexa_voice_profile_use}.

\vspace{-0.05in}
\subsection{Speaker/Speech Datasets}
\label{subsec:datasets}

We use four different speaker datasets to \htedit{define target speakers
\targ{} and their speech samples \starg{}. } 
The first three are
commonly used speaker recognition datasets, and the last one is a custom
dataset crafted for our experiments. 

\begin{packed_itemize}
\item {\bf VCTK}~\cite{vctk} contains
  short spoken phrases from 110 English speakers with varied accents. Phrases are read from a
  newspaper, the Rainbow passage~\cite{fairbanks1960voice}, and a dataset-specific paragraph. 
\item {\bf LibriSpeech}~\cite{libri} is derived from the
  open-source LibriVox audiobook project. We use the
  \texttt{test-clean} subset, which contains spoken phrases from 40 English
speakers.
\item {\bf SpeechAccent}~\cite{speechaccent} contains
  the same set of spoken phrases (in English) from each of 2140
  speakers. Speakers come from 177 different countries and represent
  214 native languages.
\item Our {\bf Custom} dataset contains
  phrases from the Rainbow passage spoken in English by 14 speakers
  (details in \S\ref{sec:blackbox}). This dataset allows us to
  synthesize speech for real-world tests on WeChat and Alexa in
  \S\ref{sec:blackbox}. 
\end{packed_itemize}

\vspace{-0.05in}
\subsection{Ethics}
All of our user study protocols received approval from our
local IRB board and were carefully designed to protect the privacy and well-being
of our participants. We kept only audio recordings of the interview sessions,
which were anonymized and stored on secure servers. Given our goal of
bringing attention to the significance of this attack vector, we have also
proactively contacted Microsoft Azure, WeChat, and Amazon to disclose our findings.

\section{Synthesized Speech vs. Machines}
\label{sec:machines}

We begin by asking ``{\em how vulnerable are machine-based SR systems to synthetic speech
attacks}?''  While prior
work has explored this question using classical (non-DNN) synthesis
systems, the efficacy of DNN synthesis attacks against real-world SR systems 
remains unknown.  In this section, we answer this question by
evaluating the robustness of four modern SR systems (\S\ref{subsec:sr})  to DNN-based 
synthesis attacks. 



Specifically, our study consists of the following experiments:
\begin{packed_itemize}
  \item As a baseline, \S\ref{sec:esorics} {\bf recreates prior classical synthesis
      attacks} and finds that they fail against newer SR systems.
  \item \S\ref{sec:whitebox} {\bf attacks the widely used SR model} 
    (Resemblyzer), showing
    that DNN-based synthesis attacks reliably fool such systems. 
  \item \S\ref{sec:azure} and \S\ref{sec:blackbox} {\bf attack
     three real-world SR deployments} (Azure,
    WeChat, and Amazon Alexa), showing that all three are vulnerable to DNN-based
    synthesis attacks.
  \end{packed_itemize}
  
\rev{We measure attack performance using the {\em attack
  success rate (AS)}, which denotes the average percent of synthesized
samples identified as the target speaker.} We design our experiments to not
only evaluate the attack success rate against various SR systems, but also explore whether a target's 
  speech samples and personal attributes (e.g., gender/accent) impact the attack outcome.  


\subsection{Baseline: Classical Synthesis Attacks against Modern SR}
\label{sec:esorics}
As reference, we evaluate the efficacy of prior classical (non-DNN) synthesis
attacks against today's  SR systems. A 2015
paper~\cite{mukhopadhyay2015all} demonstrated that synthetic speech
created with 
Festvox~\cite{festvox} fooled the Bob Spear SR system~\cite{khoury2014spear} with $>98\%$ success
rate. We recreate this attack and
find that it fails on more recent SR systems (Table~\ref{tab:festvox}).
A detailed description of our experiments is in
the Appendix.
\begin{table}[h]
\centering
\resizebox{\linewidth}{!}{%
\begin{tabular}{l|c|c|c|c} 
\hline
\multirow{2}{*}{\diagbox{\textbf{Attacker}}{\textbf{System}}} & \multicolumn{2}{c|}{\textit{Traditional SR Systems}} & \multicolumn{2}{l}{\textit{Modern SR Systems}} \\ 
\cline{2-5}
 & \begin{tabular}[c]{@{}c@{}}\textbf{Bob Spear}\\\textbf{UBM-GMM }\end{tabular} & \begin{tabular}[c]{@{}c@{}}\textbf{Bob Spear}\\\textbf{ISV }\end{tabular} & \textbf{Azure } & \textbf{Resemblyzer } \\ 
\hline
\multicolumn{1}{c|}{\textbf{ Attacker 1 (Male) }} & 90.3\% & 98.3\% & 11.1\% & 0.0\% \\ 
\hline
\multicolumn{1}{c|}{\textbf{ Attacker 2 (Female) }} & 96.0\% & 96.0\% & 2.0\% & 0.0\% \\
\hline
\end{tabular}
}
\caption{The classical synthesis attack generated by Festvox effectively fools traditional SR systems (Bob Spear)
  but fails on modern SR systems (Azure, Resemblyzer). }
\label{tab:festvox}
\vspace{-0.2in}
\end{table}
\subsection{Resemblyzer (Open Source SR)}
\label{sec:whitebox}
Next we test DNN-based speech synthesis attacks against Resemblyzer, a modern SR system
widely used in recent literature. We use the official 
implementation of Resemblyzer provided by~\cite{github_resemblyzer}. 



\para{Experiment Setup.} We seek to study the attack success rate
($AS$) of DNN synthesis attacks and its dependency 
on the target \targ's {\em
  speech sample} (\starg{}) and {\em personal attributes}.  Thus we
perform in-depth analysis on  the impact of 
(1) {\em content-specific factors}, i.e., size and quality of \starg{}
and the phonetic similarity between \starg{} and the 
synthesized speech \sadv{},  and (2) {\em
  identity-specific factors}, i.e., \targ's gender and
accent.   These are the factors that a reasonable attacker would
consider.  We list 
the corresponding experiments in
Table~\ref{tab:results_overview}. 

\begin{table}[h]
\centering
\resizebox{\linewidth}{!}{%
\begin{tabular}{c|l|l} 
\hline
\textbf{Factor} & \multicolumn{1}{c|}{\textbf{Experiment Methodology}} &
                                                              \textbf{
                                                              Attack
                                                              Success
                                                              Rate ($AS$)} \\ 
\hline
\begin{tabular}[c]{@{}c@{}}Size of \\\starg{}\end{tabular} & \begin{tabular}[c]{@{}l@{}}Generate synthetic speech, varying \\number of target samples $N$.\end{tabular} & Low~$AS$~when $N < 5$ \\ 
\hline
\begin{tabular}[c]{@{}c@{}}Quality of \\\starg{}\end{tabular}
                & \begin{tabular}[c]{@{}l@{}}Add Gaussian noise to
                    target samples \\before speech
                    synthesis.\end{tabular} & $AS = 0$ even for small noise. \\ 
\hline
\begin{tabular}[c]{@{}c@{}}Phonetic\\Similarity\end{tabular} & \begin{tabular}[c]{@{}l@{}}Vary the phonetic distance between \\target samples and synthesis output.\end{tabular} & No significant effect. \\ 
\hline
\begin{tabular}[c]{@{}c@{}}Target\\Gender\end{tabular}
                & \begin{tabular}[c]{@{}l@{}}Test male/famale target
                    speakers from\\VCTK and LibriSpeech
                    separately.\end{tabular}
                & \begin{tabular}[c]{@{}l@{}}$AS$~for female
                    targets\\$> AS$ for male targets.~\end{tabular} \\ 
\hline
\begin{tabular}[c]{@{}c@{}}Target\\Accent\end{tabular}
                & \begin{tabular}[c]{@{}l@{}}Test native/non-native
                    English speakers\\from SpeechAccent dataset
                    separately.\end{tabular}
                & \begin{tabular}[c]{@{}l@{}}$AS$ for native English
                    speakers \\$> AS$ for non-native speakers.\end{tabular} \\
\hline
\end{tabular}
}
\caption{Overview of content- and identity-specific attribute
  experiments (and results) 
  on Resemblyzer and Azure. }
\vspace{-0.25in}
\label{tab:results_overview}
\end{table}



In total,  our experiments consider $90$ target speakers randomly
chosen from three speaker datasets ($20$ randomly chosen from VCTK, $20$ from LibriSpeech
\texttt{test-clean}, and $50$ from SpeechAccent). For
each target \targ{},  we use their speech sample set \starg{} as input to
either SV2TTS or AutoVC to produce fake voices of \targ{} that contain
arbitrary speech content chosen to \emily{mimic normal conversation
  (listed in 
Table~\ref{tab:phrases} in the Appendix)}.  Since AutoVC also requires a source recording,  we choose 
the source of the same gender of the target speaker as suggested by~\cite{qian2019autovc}. \emily{For each test, we synthesize $10$
  spoken phrases per target speaker.}

We note that Resemblyzer requires a threshold to detect
whether two speech embeddings are from the same speaker.  We configure
this threshold by first enrolling the target speakers in Resemblyzer
using their real speech samples, then computing their embeddings and
choosing the threshold that minimizes Resemblyzer's equal error
rate (EER) on these speakers, using cosine similarity as the distance
metric.  When launching a synthesis attack,  the
attack is considered successful if the similarity between the attack and enrolled embeddings is
above the threshold.  For each attack, we repeat the enrollment
process 10 times (using different speech samples) and report the
average attack success rate and standard deviation.





\para{Results.} \htedit{In total, we test 13,000 synthesized speech
instances targeting 90 speakers on Resemblyzer.  The results show that SV2TTS based attack is
  highly effective against Resemblyzer, while AutoVC is ineffective. }
The size and quality of \starg{} ,  the speaker gender and accent do
impact the attack success rate,  but the phonetic similarity has minimum impact.  Next we report these results in more detail. 


{\bf \em 1) Attack success rate under the default setting:}  We start
from ``ideal'' cases where the attacker targets native English speakers
with US or British accents, and has plenty of high-quality speech
samples per target.  For this we consider target speakers from VCTK and
LibriSpeech, and configure \starg{} to include $N=10$ utterances per
target.  As such, \starg{} includes 30-120 seconds of clean audio, far more than
the amount required to run zero-shot synthesis as
claimed by~\cite{qian2019autovc, jia2018transfer} (roughly 20 seconds).  We hereby
refer to $N$ as the number of target speech samples.



\begin{figure}[t]
    \centering
    \subcaptionbox{Target speakers are native English speakers ($N$=10) \label{tab:resembylzer_baseline}}[0.5\textwidth]{\resizebox{0.5\textwidth}{!}{%
\begin{tabular}{c|cc|cc} 
\hline
\textbf{Synthesis System} & \multicolumn{2}{c|}{SV2TTS} & \multicolumn{2}{c}{AutoVC} \\ 
\hline
\textbf{Dataset} & \textit{VCTK} & \textit{LibriSpeech} & \textit{VCTK} & \textit{LibriSpeech} \\ 
\hline
\textbf{Attack Success} & $50.5 \pm 13.4\%$ & $100 \pm 0.0\%$ & $8.0 \pm 5.9\%$  & $0.0 \pm 0.0\%$ \\
\hline
\end{tabular}
}}
\hfill
    \subcaptionbox{Impact of target speech sample size $N$ \label{fig:whitebox_eval}
    }[0.5\textwidth]{
      \centering
      \includegraphics[width=0.3\textwidth]{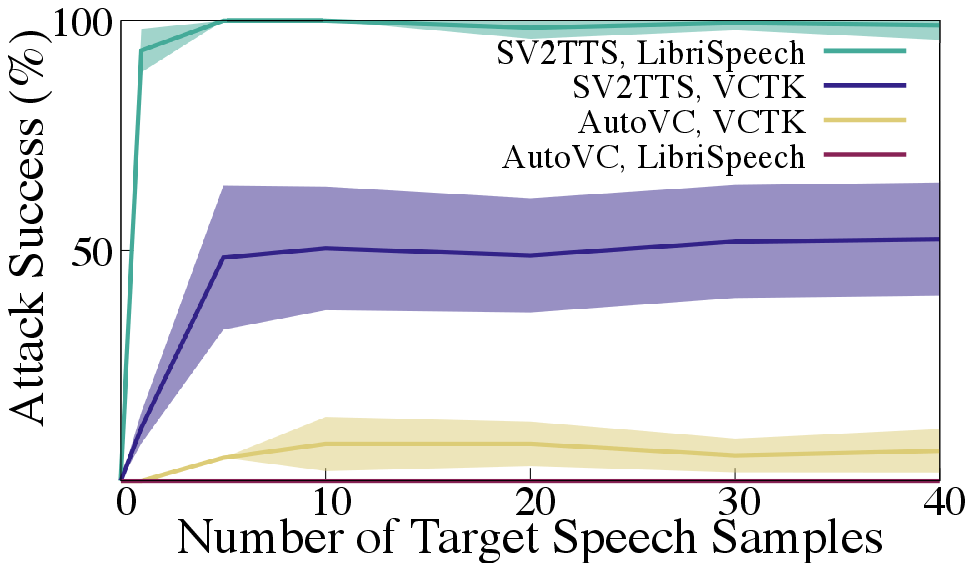}}\hfill
    \vspace{0.25cm}
    \subcaptionbox{Impact of target speaker's gender (SV2TTS, $N$=10)\label{tab:gender_resembylzer}} [0.5\textwidth]{
\resizebox{0.5\textwidth}{!}{%
\begin{tabular}{c|cc|cc} 
\hline
\textbf{Dataset} & \multicolumn{2}{c|}{VCTK} & \multicolumn{2}{c}{LibriSpeech} \\ 
\hline
\textbf{Speaker Gender} & \textit{Male} & \textit{Female} & \textit{Male} & \textit{Female} \\ 
\hline
\textbf{Attack Success} & $43.3 \pm 16.1\%$ & $61.8 \pm 9.4\%$ &
                                                                  $100.0
                                                                  \pm
                                                                  0.0\%$
                                        & $100.0 \pm 0.0\%$  \\
\hline
\end{tabular}
}}   \vspace{-0.15in}
   \caption{Attack success rate when testing DNN-based synthesize
     attacks against Resemblyzer.}
   \label{fig:resemblyzer_results}
   \vspace{-0.15in}
\end{figure}

As shown in Figure~\ref{tab:resembylzer_baseline},  fake speech
synthesized by SV2TTS successfully fools Resemblyzer, while AutoVC
fails. We think that the success of SV2TTS (particularly on LibriSpeech) is
likely because Resemblyer is trained using 
the same loss function used by SV2TTS' speaker
encoder~\cite{wan2018generalized}. 



{\bf \em 2) Impact of \starg{}  size:}  We repeat the above experiment
but vary the size of target speech samples, i.e., $N$=1, 5, 10, 20,
30, 40 speech samples. 
As Figure~\ref{fig:whitebox_eval} shows, the attack
success rate for SV2TTS grows with $N$ but levels off before $N$
reaches 10. For AutoVC, the attack remains ineffective when
varying $N$. 

{\bf \em 3) Impact of \starg{}  quality:}
This
question has bearing on real-world attack settings, since an attacker
might not always obtain high-quality audio recordings of the target.
To emulate low-quality data, we add four different levels of zero-mean
Gaussian
noise to the original clean 
audio. We vary the signal-to-noise ratio from $4$ dB (noise quieter 
than the speaker's voice) to $-15$ dB (noise louder than the speaker's voice).
We find that noisy target speech samples decimate synthesis attack performance. 
For both SV2TTS and AutoVC, the attack success rate reduces to $0\%$
at all four noise levels.

{\bf \em 4) Impact of phonetic similarity between \starg{} and
  \sadv{}:}  This factor also has strong real-world implications -- if the
content similarity affects the attack success rate, the attacker may
be largely constrained by which \starg{} they obtain. Since SV2TTS generates synthesized speech from arbitrary text,
we use it to explore this question. Details of
our experiments are in the Appendix.


Interestingly,  we find that phoneme similarity of \starg{} and
  \sadv{} does not have any 
visible 
effect -- the attack success rate remains stable as we vary
the normalized phoneme similarity from $0$ to $1$.

{\bf \em 5) Impact of target gender:} We now consider a target's personal
attributes which may affect the attack outcome. The first is speaker
gender, which can come into play if, for example, synthesis or SR models lack
sufficient gender diversity. To study this factor,  we separate the results of our
SV2TTS experiments by gender. 
We find that synthesized female speakers have higher $AS$ on average than male
speakers (Figure~\ref{tab:gender_resembylzer}). When we test clean (non-synthesized) speech from these target
speakers on Resemblyzer, the SR accuracy is 100\% for both male and
female speakers.




{\bf \em 6) Impact of target accent:} 
Most public speech datasets consist of native English
speakers with US or UK accents (i.e. VCTK, LibriSpeech, VoxCeleb 1/2). Speech synthesis systems trained on these
datasets may fail to recreate the unique prosody of speakers with different accents.
To test this, we choose 50 target speakers from the SpeechAccent
dataset, including male/female native English speakers and male/female
native speakers from the top 21 most spoken languages. 

When comparing results from native/non-native English
speakers, we observe a higher attack success rate among native
English speakers (100\%) compared to non-native English speakers
(65\%) for synthesized speech produced by SV2TTS. As before,
attacks using synthesized speech from AutoVC are unsuccessful. 

\vspace{-0.1in}
\subsection{Azure (Open API, Real-World SR)}
\label{sec:azure}


We run the same experiments of 
\S\ref{sec:whitebox} on Azure, a real-world SR deployment.  Azure's open API allows us
to enroll speakers and run numerous tests against their
enrolled speaker profiles.  But unlike 
\S\ref{sec:whitebox},  there is no need to configure any 
threshold. \emily{We enroll each of our $90$ target speakers from~\ref{sec:whitebox}
into Azure and use these enrolled profiles for all tests. We generate
and test $10$ synthesized phrases (as in~\ref{sec:whitebox}) per target for each experiment. Since
Azure reports the SR acceptance result per sample, we report the
average success rate of all synthesized samples in each experiment.}



\para{Disclosure.} We followed standard disclosure practices and
reported the result of DNN-synthesized speech attacks to Microsoft.


\para{Results.} \htedit{We test 13,000 synthesized speech
instances targeting 90 speakers on Azure}.  These results show that Azure is also
vulnerable to  DNN-synthesized speech.   Our findings on the impact of various factors mirror those from
Resemblyzer. 

\begin{figure}[t]
    \centering
    \subcaptionbox{When target speakers
are native English speakers ($N$=10)\label{tab:baseline_azure}}[0.5\textwidth]{\resizebox{0.5\textwidth}{!}{%
\begin{tabular}{c|cc|cc} 
\hline
\textbf{Synthesis System} & \multicolumn{2}{c|}{SV2TTS} & \multicolumn{2}{c}{AutoVC} \\ 
\hline
\textbf{Speaker Source} & \textit{VCTK} & \textit{LibriSpeech} & \textit{VCTK} & \textit{LibriSpeech} \\ 
\hline
\textbf{Attack Success} & $29.5 \pm 32.0\%$ & $21.5 \pm 23.5\%$ & $14.5 \pm 34.1\%$  & $0.0 \pm 0.0\%$ \\
\hline
\end{tabular}
}}
\hfill
    \subcaptionbox{Impact of target speech sample size $N$ ($\sigma$ omitted for clarity).\label{fig:azure_eval}
    }[0.5\textwidth]{
      \centering
      \includegraphics[width=0.3\textwidth]{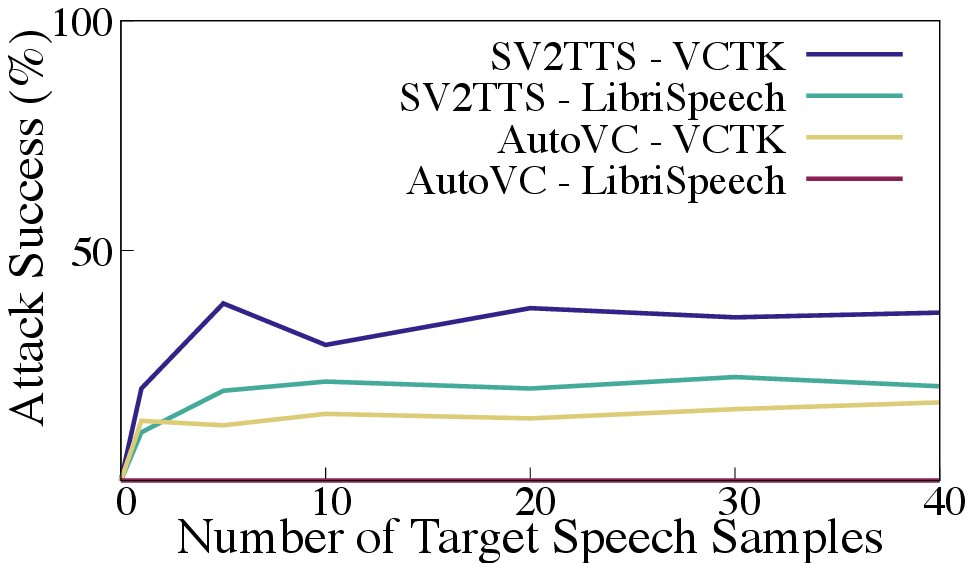}}\hfill
    \vspace{0.25cm}
    \subcaptionbox{Impact of target speaker's gender (SV2TTS, $N$=10) \label{tab:gender_azure}} [0.5\textwidth]{
\resizebox{0.5\textwidth}{!}{%
\begin{tabular}{c|cc|cc} 
\hline
\textbf{Speaker Source} & \multicolumn{2}{c|}{VCTK} & \multicolumn{2}{c}{LibriSpeech} \\ 
\hline
\textbf{Speaker Gender} & \textit{Male} & \textit{Female} & \textit{Male} & \textit{Female} \\ 
\hline
\textbf{Attack Success} & $7.8 \pm 13.9\%$ & $47.3 \pm 32.0\%$ & $14.0 \pm 22.2\%$ & $29.0 \pm 23.3\%$ \\
\hline
\end{tabular}
}}\vspace{-0.15in}
   \caption{Attack success rate when testing DNN-based synthesize
     attacks against Azure.}\vspace{-0.15in}
    \label{fig:azure_results}
\end{figure}

{\bf \em 1) Attack success rate under default settings:}
Figure~\ref{tab:baseline_azure} lists the overall attack success
rate.  We see that DNN-synthesized speech can easily fool Azure, although the attack
success rate is less than those with Resemblyzer.  Interestingly, for
62.5\% of target speakers, at least 1 out of 10 synthesized phrases
(generated by SV2TTS)  was
accepted by Azure as the target speaker.  Thus a persistent attacker
could make multiple attempts to eventually fool Azure API (assuming there is no
limit on authentication attempts). 

Another interesting finding is that the attack
success rate displays significantly
higher variance than those observed on Resemblyzer.  This is
particularly visible for SV2TTS. When we dig deeper to understand this
high variance, we find that for the above mentioned 62.5\% targets
(with at least 1 success attack instance out of 10 trials), the attack success rate was
  $49.2 \pm 23.5\%$ for VCTK speakers and
$33.1 \pm 21.4\%$ for LibriSpeech speakers. These results indicate
that the 
attack performance against Azure is non-uniform across target
speakers.


{\bf \em 2) Impact  of  \starg{} size, quality, and phonetic
  similarity of \starg{} and \sadv:}  Our results from these
experiments mirror those of Resemblyzer:  Figure~\ref{fig:azure_eval}
shows that the attack performance levels off when $N$ reaches 10;  none
of the speech synthesized from noisy versions of \starg{} was accepted
by Azure; and the phonetic
  similarity of \starg{} and \sadv{} does not affect the attack
  outcome.



{\bf \em 3) Impact of target gender:}  We 
observe a significant difference in the attack success rate for male and female
targets enrolled in Azure. Figure~\ref{tab:gender_azure} reports the
results using SV2TTS, where attacks targeting female speakers are
much 
more effective. Similar trends are observed on AutoVC (VCTK). Again,
when we test Azure with clean (non-synthesized) speech of our target
speakers,  the SR accuracy is 100\% for both male and female
speakers. 





{\bf \em 4) Impact of target accent:}  We find that SV2TTS-synthesized
samples for  SpeechAccent speakers lead to an
attack success rate of  
$8.8 \pm 14.5\%$. Among them, the attack success rate is $15.0 \pm
16.9\%$ for native English speakers but drops to $7.5 \pm 13.9\%$
for non-native English speakers with accents.


\subsection{WeChat and Amazon Alexa (Closed API, Real-World SR)}
\label{sec:blackbox}

Finally, we experiment with two additional real-world SR systems: WeChat and
Amazon Alexa. In contrast to Azure, both employ closed-API SR, largely limiting our experimental bandwidth. Since
WeChat and Alexa's SR systems link to individual
accounts, we must test synthesis attacks with real users. \emily{Note that
  our goal is not to test the (in)security of WeChat or Alexa platforms, but
  to use them as case studies of deployed SR systems to
  illustrate the potential impact of DNN-based speech synthesis attacks.}

\para{Experimental Setup.}  We conduct an IRB-approved  user study to evaluate synthetic speech attacks (IRB information omitted
for anonymity).  Specifically,  we collect speech samples from study participants, synthesize speech imitating each participant, and give 
these speech samples to each participant to test their WeChat and Amazon
Alexa apps. Given the poor performance
of AutoVC on Resembylzer and Azure, we only use SV2TTS for these
experiments. 

We recruited $14$ participants with different 
linguistic background (1 native Marathi speaker; 1 native Dutch
speaker; 3 native Mandarin speakers; 9 native English speakers) and gender (10
female/4 male). All participants signed written consent forms to
participate in our user study and were compensated \$10 for their
time.  We asked our  participants to submit a small set of their voice recordings. 
Each participant used a voice memo recording
app to record themselves speaking 20 sentences in English from the Rainbow
Passage. The Rainbow Passage is commonly used in linguistic studies
because it contains most of the phoneme combinations in the English
language~\cite{fairbanks1960voice} (the full passage is in the
Appendix).  In this study, 7 participants recorded on a Macbook Pro, 4
recorded on an iPhone 11+ phone, and 1 recorded on a Google
  Pixel phone. 


For each participant \targ{}, we use their submitted speech recordings as the target speech
sample set \starg{}, and input them to SV2TTS to
generate synthetic speech imitating \targ{}. 
The content of the synthetic speech \sadv{} is designed to match the context of the SR system,
which we describe below (also see Table~\ref{tab:alexa}). 

\begin{packed_itemize}\vspace{-0.05in}
\item {\bf WeChat} uses a {\em text-dependent} speaker verification
  system that asks for stating the same eight-digit login number for each SR
  attempt. Each participant consented to share their unique login
  number with our user study administrators,
  and these were used to generate synthesized login speech samples. To ensure
  participant privacy and security, login numbers were password-protected, anonymized, and deleted
  when the study ended. For each participant, we generate six synthesized login samples.
\item {\bf Alexa} employs a {\em text-independent} speaker
  verification system, but its specific uses of voice profiles
  constrain the samples we can test.  We create a short list of Alexa
  commands that Amazon explicitly states should be linked to a user's voice profile
  , restricting our attention to
  native Alexa skills~\cite{alexa_voice} (see Table~\ref{tab:alexa}). \vspace{-0.05in}
\end{packed_itemize}

After setting up their WeChat voice login and Alexa voice profile (using the Alexa
smartphone app), all 14 participants verified that they could use their real voices to log into WeChat and access specified
applications with Alexa. They were then given their
synthesized speech samples (6 samples of login for WeChat, 7 samples
of commands for Alexa) and instructed to play each sample over a computer loudspeaker
located six inches from their phone microphone. Samples were played as the
targeted apps were set up to perform normal voice authentication. Each participant tested the WeChat
samples twice and the Alexa samples once. Participants recorded how the apps
responded to the synthesized samples and reported their results via a
standardized form.

\para{Attack Evaluation.} In total, our user study tested
WeChat and Alexa with 168 and 98 synthesized speech instances, respectively. 
Again, we use {\em attack success rate} ($AS$) to
evaluate how effectively synthesized speech can fool both SR
systems. For WeChat,  each attack instance is successful if the login
is approved. For Alexa,  we use a different approach because Alexa
does not provide clear-cut success/failure results:  a synthesis
attack 
instance \sadv{} succeeds if Alexa responds to \sadv{} the
same way it responds to a clean (non-synthesized) version of the
command.

\begin{table}[t]
\centering

\resizebox{0.95\linewidth}{!}{
\begin{tabular}{c|l|r} 
\hline
\textbf{SR System} & \multicolumn{1}{c|}{\textbf{Phrases Used for SR}}                     & \multicolumn{1}{c}{\textbf{$AS$}}  \\ 
  \hline
  WeChat    &  < The user states the 8-digit login number>                        & $64.3\%$                                      \\
\hline

          & Hey Alexa add an event to my calendar for tomorrow at 5.   & $71.4\%$                                      \\ 
\cline{2-3}
          & Hey Alexa check my email                                   & $64.3\%$                                      \\ 
\cline{2-3}
 Amazon         & Alexa say who is talking with you now                      & $35.7\%$                                      \\ 
\cline{2-3}
Alexa     & Alexa tell me what is on my calendar                       & $50.0\%$                                      \\ 
\cline{2-3}
          & Tell me what is on my calendar for this week               & $85.7\%$                                      \\ 
\cline{2-3}
          & Alexa make an appointment with my doctor                   & $57.1\%$                                      \\ 
\cline{2-3}
          & Hey Alexa make a donation to the American Cancer Institute & $71.4\%$                                      \\ 
\cline{2-3}
          & <\textit{Average across the above 7 commands}>                               & $62.2\%$                                      \\ 
\hline
\end{tabular}}
\caption{The phrases used in our Amazon Alexa and WeChat experiments
  and the corresponding attack success rate ($AS$) on synthesized versions of each phrase.}
\label{tab:alexa}
\vspace{-0.35in}
\end{table}

\para{Results.} On average, our speech synthesis attacks had a $63\%$
$AS$ across all tests on WeChat and Alexa. 

{\bf WeChat:} 9 of the 14 participants ($64\%$) successfully logged into their WeChat account using a synthesized
speech sample. In general, this indicates that speech synthesis attacks are a viable
authentication attack against WeChat. 
However, the number of successful fake login 
samples varied significantly across participants (despite using the
same setup for each test). On average, $1.33 \pm 1.67$ fake login
samples succeeded per participant. For one participant, all six login
samples worked. For 8 other participants who successful logged in,
only one or two
samples consistently worked.

{\bf Alexa:} Our attacks on Alexa were similarly successful ($62.2\%$
$AS$ on average)
. All 14 participants had at least 2 synthesized commands that fooled Alexa. These synthesized commands were able to access private emails, check calendar
appointments, and request financial transactions. Table~\ref{tab:alexa} reports
these results.  Furthermore, in limited tests, we found that
sometimes a wrong voice (i.e. from a different person) was
able to access user data purportedly protected by voice profiles.

{\bf Effect of Loudspeaker:} \htrev{Our study participants reported the
  device used to play their attack samples. Devices used include LG
  desktop monitors, Macbook Pros, Bose Soundlink Speakers, and iPhone 11s. We examined the
  impact of speaker hardware on attack success and found no
  correlation between the two.  While the tested devices already cover a
  broad set of speaker hardwares found in today's households and offices, 
  more experiments are needed to quantify if attack
  success depends on speaker quality.}


\para{Disclosure.} We followed standard disclosure practices and
reported our attacks to both WeChat and Amazon. 


\rev{
\subsection{Key Takeaways}
\label{sec:systems_discussion}

\htrev{All four modern SR systems tested are vulnerable to DNN-based speech
  synthesis attacks, especially those generated by SV2TTS. It is
  alarming that for three popular real-world SR systems (Azure, WeChat, Alexa),
  more than $60\%$ of enrolled speakers have at least $1$ synthesized
  (attack) sample accepted by these systems. This clearly demonstrates the
  real-world threat of speech synthesis attacks. 

  Another key observation is that the attack performance is
  speaker-dependent, e.g.,  the number of synthesized
  samples that successfully fooled the SR systems varies across speakers.  For Resemblyzer and Azure, the attack success rate is
  consistently higher for female and native English speakers.

  \para{Limitations and Next Steps.}   Our experiments,
  especially those on WeChat and Alexa, involved a moderately-sized set of target
  speakers to demonstrate the real-world threat of speech
  synthesis attacks. To further evaluate the attack dependence on
  target human speakers, we believe viable next steps include expanding the
  speaker pool and testing more operational scenarios. With these two
  changes, we could more closely examine how an individual's vocal
  characteristics (e.g., pitch, accent, tone) affect the attack success rate, 
  and whether their impact can be reduced by improving the
  underlying speech synthesis systems. 

Similarly, due to our focus on low-resource attackers, our experiments used two publicly available
  speech synthesis systems (SV2TTS and AutoVC) that were trained only
  on publicly available datasets. These two systems will likely underperform advanced synthesis
systems trained on larger, proprietary datasets, and consequently our reported results only offer a
``conservative'' measure of the threat.  As speech synthesis systems
continue to advance, the threat (and damage) of speech
synthesis attacks will grow and warrant our continuous attention.}

}


\vspace{-0.08in}
\section{Synthesized Speech vs. Humans} 
\label{sec:humans}
\htedit{Having demonstrated that DNN-synthesized
  speech can easily fool machines (e.g., real-world SR systems), we
  now move to evaluate their impact on humans.  Different from prior work
  that uses surveys to measure human perception of
speech synthesized by classical (non-DNN)
tools~\cite{mukhopadhyay2015all, neupane2019crux}, we assess
the susceptibility of humans to DNN-synthesized speech in different
interactive settings. For this
we conduct two user studies, covering both static survey and
``trusted'' interaction settings.  Next, we describe
the methodology behind our user studies and give a preview of our key
findings, before presenting both studies in detail. 
}




\vspace{-0.1in}
\subsection{Methodology and Key Findings} 
An attacker \adv{} can perform a variety of attacks against
human listeners using synthesized speech. Such attacks can be particularly
effective if the listener has limited familiarity with the owner of
the spoofed voice.  For example, \adv{} could use a synthesized voice to
perform the classic spear-phishing attack, where an elderly victim gets a
phone call from their ``grandchild'' they haven't seen in months who
is stuck in a foreign country and needs emergency cash to get home, or an employee gets a call from their ``boss''
confirming an earlier (phishing) email authorizing a money transfer~\cite{fakevoice_theft}.

With these in mind, our study on the impact of synthesized voices on human
listeners has two goals: understanding human listeners' susceptibility to
synthesized speech in isolation and in ``trusted contexts.'' We designed
experimental protocols for both parts of our study, giving detailed
consideration to issues of ethics and impact on our participants. All
protocols were carefully evaluated and approved by our institutional IRB
review board. We discuss ethical considerations in~\S\ref{sec:discussion}.

\para{User Study A (Online Survey).} We first evaluate {\em whether}  humans can
discern the difference between real and DNN-synthesized (fake) speech. We
conduct an online user survey and compare how well participants could identify
synthesized speech for voices with which they have varying levels of familiarity (\eg strangers
vs. celebrities).
We also measure the effect of priming by comparing
results from 2 scenarios: one in which participants are told the speech
samples will contain a mix of real and fake speech, and one without the
disclosure.

{\bf \em Finding:} In this ``survey'' setting, DNN-synthesized speech
fails to consistently fool humans. Participants could more
easily distinguish between real and fake speech
when they were more familiar with
the speaker, and when they were aware that some speech may be fake
(thus tended to listen carefully with added skepticism). 

\para{User Study B (Deceptive Zoom Interview).} We seek to better understand the impact of {\em context} on
listeners' susceptibility to fake speech. To do so, we conduct
interviews over Zoom calls.  Participants believed they were speaking
to two (human) researchers, but in reality one of the voices was
synthesized speech. 

{\bf \em Finding:} In this ``trusted setting,'' all 14 participants showed no hesitancy or
suspicion during the interview, and readily responded to and complied with
all requests from the ``fake interviewer.''  In other words, a synthesized
voice consistently fooled humans in this trusted interview setting. 



\subsection{User Study A:   Can Users Distinguish Synthesized and Real Speech?} 
\label{sec:survey}

We begin our human perception experiments with the critical question: ``can 
human listeners distinguish synthetic speech of a speaker from the real thing?'' 
We deploy a survey to assess users' ability to
distinguish between real and fake speakers. 


\para{Participants.} 
 We recruited 200 participants via
 the online crowd source platform Prolific
 (\url{https://www.prolific.co/}).
 \htedit{All self-identified as native English
  speakers residing in the United States.}
Of our participants, 57\% identified as female (43\% male).
The participants \emily{are all $18+$ years old and} cover multiple age groups: 18-29 (43\%),
30-39 (32\%), 40-49 (14\%), 50-59 (8\%), 60+(3\%).
The survey was designed to take 10 minutes on average,
and participants received \$2 as compensation.
The study was approved by our local IRB.

\para{Procedure.}  The participants completed an online survey consisting of
several speech samples presented in pairs for side-by-side comparison.  Each
pair of samples contains one of the three following combinations: two real speech
samples of the same speaker (referred to as ``Real A/Real A'' in this section);
one real speech sample from a speaker and one real speech sample from a
different speaker (``Real A/Real B''); or one real speech sample from a
speaker and one fake speech sample imitating the speaker (``Real A/Fake
A''). We generate fake speech using SV2TTS, using ~30 seconds of clean
speech samples \starg{} from the speaker. 


{\bf Types of Speakers: } We included speakers whose
(real) voices have varying levels of familiarity with our participants:

\begin{packed_itemize}
  \item {\em Unfamiliar speakers: } Speakers from the VCTK~\cite{vctk}
    dataset whose voices have (most likely) never been heard by
    the participants.
  \item {\em Briefly familiar speakers: } Inspired
    by~\cite{mukhopadhyay2015all}, we included a set of speakers whose
    voices the participants hear only briefly. For each speaker, we provided
    participants with a short audio clip to familiarize them with the
    speaker's voice. There are four briefly familiar speakers, and for
    each one we provided a different length audio clip -- 30 seconds
    for the first speaker, 60 seconds for the second, 90 seconds for
    the third, and 120 seconds for the fourth.
  \item  {\em Famous speakers: } We used the voices of two
    American \jenna{public figures}:
    Donald Trump and Michelle Obama. We asked 
    participants whether they have heard these voices outside the
    context of this survey, and over $90\%$ responded ``yes.''
  \end{packed_itemize}

\para{Task.}  Participants listened to pairs of speech samples and
reported if both samples were spoken by the same person. 
 

  \para{Conditions.}  We deploy two versions of the survey.  Both versions
  ask participants to assess the identity of the speaker and quality
  of speech samples.  The first version does not mention fake speech at all.
  The second version of the survey mentions fake speech, both in its title
  and in its description of the task.

\begin{table}[t]
\centering
\resizebox{0.49\textwidth}{!}{%
\begin{tabular}{cc|c|c|c|c|c|c|c|c}
  \cline{2-10}
\multirow{2}{*}{} &
  \multicolumn{3}{c|}{\textbf{\begin{tabular}[c]{@{}c@{}}Unfamiliar\end{tabular}}} &
  \multicolumn{3}{c|}{\textbf{\begin{tabular}[c]{@{}c@{}}Briefly Familiar \end{tabular}}} &
  \multicolumn{3}{c}{\textbf{\begin{tabular}[c]{@{}c@{}}Famous \end{tabular}}} \\ \cline{2-10} 
 &
  \multicolumn{1}{c}{\textit{Yes}} &
  \multicolumn{1}{c}{\textit{\begin{tabular}[c]{@{}c@{}}Not \\ Sure\end{tabular}}} &
  \multicolumn{1}{c|}{\textit{No}} &
  \multicolumn{1}{c}{\textit{Yes}} &
  \multicolumn{1}{c}{\textit{\begin{tabular}[c]{@{}c@{}}Not\\ Sure\end{tabular}}} &
  \multicolumn{1}{c|}{\textit{No}} &
  \multicolumn{1}{c}{\textit{Yes}} &
  \multicolumn{1}{c}{\textit{\begin{tabular}[c]{@{}c@{}}Not\\ Sure\end{tabular}}} &
  \multicolumn{1}{c}{\textit{No}} \\ \hline
\multicolumn{1}{c|}{\textbf{Real A / Real A}}      
& $\mathbf{90.9\%}$ & $9.1\%$ & $0\%$ & $\mathbf{69.6\%}$ & $17.5\%$ & $12.9\%$  & $\mathbf{76.4\%}$ & $16.7\%$ & $6.9\%$ \\ \hline
\multicolumn{1}{c|}{\textbf{Real A / Real B}} 
& $0\%$ & $6.7\%$ & $\mathbf{93.7\%}$ & $0\%$ & $3.3\%$ & $\mathbf{96.7\%}$ & $0\%$ & $9.5\%$ & $\mathbf{90.5\%}$  \\ \hline
\multicolumn{1}{c|}{\textbf{Real A / Fake A}} 
& $17.3\%$ & $32.7\%$ & $\mathbf{50.0\%}$ & $18.5\%$ & $31.2\%$ & $\mathbf{50.3\%}$ & $4.1\%$  & $16.0\%$ & $\mathbf{79.9\%}$  \\ \hline
\end{tabular}%
}
\caption{Participants' answers when asked if  two voice samples
were from the same person. We use this to gauge their ability to correctly discern if
  speech samples were spoken by the same speaker (Real A/Real A),
  different person (Real A/Real B), or
  a synthesized (fake) speaker (Real A/Fake A). }
\vspace{-0.1in}
\label{tab:study_similarity}
\end{table}


%

\para{Results.}  We seek to answer the following questions:

{\bf \em 1) Do participants think generated fake speech 
    was spoken by the original speaker?}
  As shown in Table~\ref{tab:study_similarity} (bottom row), 
   about half of participants were fooled, {\em i.e.} they
  responded ``yes'' or ``not sure,'' when asked this question about
  {\em unfamiliar} or {\em briefly familiar} speakers. 
  For {\em famous} speakers whose voice participants are generally familiar
  with, this number drops to $20\%$. 

  {\bf \em 2) Does hearing more samples from a speaker (\ie knowing a
    speaker better) make fake speech more detectable?}
    Results in Table~\ref{tab:study_similarity} suggest that greater familiarity with a
  speaker will lead to increased skepticism of a fake voice. 
  Compared to a similar user study performed 6 years
  ago~\cite{mukhopadhyay2015all}, proportion of participants who
    correctly identified the fake voice for unfamiliar or briefly familiar
    speakers is consistent ($50\%$ for our work vs. $48\%$
    in~\cite{mukhopadhyay2015all}). However, participants in our survey were
    more accurate at identifying fake speech from famous speakers (
    $80\%$ vs. ~$50\%$ in~\cite{mukhopadhyay2015all}), perhaps reflecting a
    higher general awareness of speech synthesis attacks. 
  
  \begin{figure}[t]
    \vspace{-0.3in}
    \centering
    \includegraphics[width=0.48\textwidth]{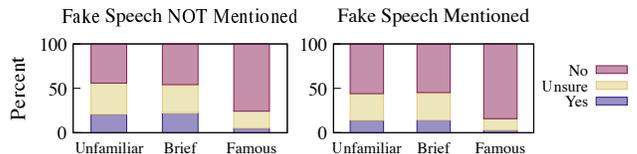}
       \vspace{-0.3in}
    \caption{User responses to the question ``are these two voice
      samples from the same person?''
      (Left) when users are not told
      synthesized speech is used in the survey; (Right) when users are told this.}   \vspace{-0.1in}
    \label{fig:nofake_fake}
  \end{figure}

{\bf \em 3) Does mentioning fake speech in the survey description change
      participants' perceptions of the fake speech samples?}
Mentioning fake speech in the survey description
showed a statistically significant effect on survey
responses. Figure~\ref{fig:nofake_fake} shows how the responses to the
survey version mentioning fake speech 
reflect an apparent increased 
skepticism of fake speakers. 

Using a chi-squared test for independence, we compared responses from each
speaker familiarity category 
between the two survey versions to see if this change is statistically
significant. 

\begin{packed_itemize}
\item For {\em unfamiliar} speakers, all but one speaker has a
  significant ($p < 0.05$) difference in responses.
  \item For {\em somewhat familiar} speakers, again all but one
    speaker has a significant ($p < 0.05$) difference in responses.
  \item For {\em famous} speakers, only Trump has a statistically
    significant difference in responses. 
  \end{packed_itemize}

  \rev{
{\bf \em 4) Do participant demographics (age, gender) affect
  responses?} Women and younger people were more likely to correctly identify
real and fake speakers. Using a chi-squared
test for independence, we compared the responses of men vs. women and
younger people (age < 25) to older people (age > 45). For {\em
  unfamiliar speakers}, there were statistically significant ($p <
0.05$) differences in response between genders and age groups. For
somewhat familiar and famous speakers, statistically significant
differences are observed for some, but not all, speakers.
    }

\subsection{User Study B: How Do Users Interact with Synthesized  
  Speech in Trusted Settings?}
\label{sec:trusted}


Our user study A confirms that DNN-synthesized speech 
fails to {\em consistently} fool humans in a survey setting.  \jenna{Looking
  beyond this, we wonder how, if at all, the {\em context} in which users were exposed to fake speech influences
their susceptibility to these attacks.}
Specifically, how would participants \emily{behave} in a setting where they were
predisposed {\em not} to think critically about the voices they hear?
Examples of such ``trusted settings'' include phone or Zoom meetings with colleagues
or calls with one or more people they know (or think they know). 
Human behavior in these so-called ``trusted settings'' may differ from
behavior in a survey-based setting. 
When humans are primed by their setting to think they are
speaking to a real person, they may be more likely to accept fake
speech as real.

\para{Study Design.} 
To understand the impact of trusted settings on human interactions
with fake speech, we conduct a user study involving deceptive
interviews.

{\bf{\em Ethics: }} This study was approved by our institutional IRB.  Participants
submitted a signed consent form prior to the interview and received a full
debriefing afterwards to inform them of the deception and true purpose of the
study.  No personal information about participants was retained after the
interview, and interview recordings were anonymized to protect
participant privacy.

\para{Participants.}
Interviewees were recruited from among the 
students in our institution's computer science
department. We conducted a total of 14
interviews. Twelve interviewees were male (2 female). All were
between the ages of 20 and 35 years old, with varied ethnic/racial backgrounds
(American, Chinese, Indian, Indonesian, Turkish). 
The interviews were approximately 10 minutes,
and participants were compensated with a \$10 Amazon gift card.

\para{Procedure.}
The recruitment call 
asked for participation in an interview study about use of speech
recognition systems (\eg Siri) and their perceptions of privacy with
respect to these systems.
Each interview took place over a Zoom call, with two
paper authors functioning as ``interviewers.'' One of the interviewers
(hereafter referred to as the {\em real interviewer}) used their real
voice throughout the staged interview, while the other (referred to as the {\em fake interviewer})
used only fake speech samples based on their real voice. All fake
speech samples were generated using the SV2TTS 
synthesis system and less than 5 minutes of real voice samples from
the fake interviewer. Throughout the call, the fake speech
samples are played from an iPad Pro, held close to the fake interviewer's computer microphone.

\rev{
After the conclusion of the deceptive portion, we revealed the use of fake 
voice samples and disclosed our research objectives
before asking a few additional questions. Participant responses were categorized and coded
separately by each interviewer, who later met to combine the codes and
resolve discrepancies. All themes in responses described below were
expressed by $>=3$ participants, unless otherwise noted. More details
about the post-deception interview and analysis procedure can be found
in the Appendix.
}   

Because all of the interviewees were members of the
authors' academic division, they had different levels of familiarity with the
interviewers, ranging from general knowledge to frequent social
interaction. At the end of each interview, participants were asked to rank 
their familiarity with the interviewers' voices prior to the interview on
a scale of 1 (``not at all familiar'') to 5 (``extremely
familiar''). Table~\ref{tab:familiarity} lists the distribution of 
familiarity rankings. 

\begin{table}[h]\vspace{-0.05in}
\centering
\resizebox{\linewidth}{!}{%
\begin{tabular}{cccccc} 
 & \multicolumn{1}{c}{\begin{tabular}[c]{@{}c@{}}\textit{Not at all}\\\textit{Familiar}\end{tabular}} & \multicolumn{1}{c}{\begin{tabular}[c]{@{}c@{}}\textit{Slightly}\\\textit{Familiar}\end{tabular}} & \multicolumn{1}{c}{\begin{tabular}[c]{@{}c@{}}\textit{Moderately }\\\textit{Familiar}\end{tabular}} & \multicolumn{1}{c}{\begin{tabular}[c]{@{}c@{}}\textit{Very }\\\textit{Familiar}\end{tabular}} & \multicolumn{1}{c}{\begin{tabular}[c]{@{}c@{}}\textit{Extremely }\\\textit{Familiar}\end{tabular}} \\ 
\hline
\multicolumn{1}{r|}{\textbf{Real Interviewer}} & 9  & 1 & 2 & 0 & 2 \\ 
\hline
\multicolumn{1}{r|}{\textbf{Fake Interviewer}} & 7  & 2 & 3 & 1 & 1 \\
\hline
\end{tabular}
}
\caption{\# of participants and their declared familiarity with the
  two interviewer's voices before the Zoom interview.} \vspace{-0.25in}
\label{tab:familiarity}
\end{table}

\para{Task.}
The staged interview itself consists of 8 questions 
\jenna{about use of automatic speech recognition systems and perceptions of privacy} 
(see Table~\ref{tab:questions}). Five are asked by the real
  interviewer, and three are asked by the fake interviewer. The
three fake interviewer questions are designed to solicit three
different types of behavior from participants: conversational response
(Q2), website access (Q5), and personal information (Q7).

\begin{table}[h]
\centering
\resizebox{\linewidth}{!}{%
\begin{tabular}{c|c|l}
\textbf{\#} & 
\textbf{Interviewer} & \multicolumn{1}{c}{\textbf{Question}} \\ 
\hline
1 & 
Real & \begin{tabular}[c]{@{}l@{}}Do you use automatic speech recognition systems in \\everyday life?\end{tabular} \\ 
\hline
2 & 
\textit{Fake} & \textit{How often do you use these systems in your daily life?} \\ 
\hline
3 & 
Real & What do you do in your interactions with these systems? \\ 
\hline
4 & 
Real & \begin{tabular}[c]{@{}l@{}}Do you ever think about your privacy during your \\interactions with these systems?\end{tabular} \\ 
\hline
5 & 
\textit{Fake} & \textit{Can you visit this website? I'll put the link in the chat.} \\ 
\hline
6 & 
Real & \begin{tabular}[c]{@{}l@{}}Have you ever used the ``voice profiles'' feature of these \\systems?\end{tabular} \\ 
\hline
7 & 
Real & \begin{tabular}[c]{@{}l@{}}Are you ever concerned about privacy if/when you use \\voice profiles?\end{tabular} \\ 
\hline
8 & 
\textit{Fake} & \begin{tabular}[c]{@{}l@{}}\textit{We need your student id to track your participation in this }\\\textit{study. Can you leave it in the chat?}\end{tabular} \\
\hline
\end{tabular}
}
\caption{Questions asked by real and {\em fake} interviewers.}
\vspace{-0.25in}
\label{tab:questions}
\end{table}

\para{Conditions.}
Participants are not told that the 
\jenna{study is actually about perceptions of fake speech, }
and they do not know that one of the interviewers is using
a fake voice. When the participant joins the Zoom call, the real
  interviewer informs them that everyone in the call is keeping their
video off to preserve interviewee privacy. In reality, keeping videos
off prevents the participant from observing that the fake interviewer is using a fake voice. 
\jenna{We also asked the interviewees if we could record the interview 
to maintain a record of their responses.}

Because of the relatively low quality of the fake interviewer voice (see
\S\ref{sec:survey}), interviewees are primed to expect a low quality
voice from the fake interviewer. 
\jenna{For 10 of the 14 participants, before beginning the interview questions,} 
the real interviewer notes
that the fake interviewer is feeling unwell and will only chime
in intermittently during the interview.
We examine the effect of excluding this {\em priming  statement} from the
interview in later sections.

\para{Results.} 
{\em None of the participants exhibited any suspicion
or hesitancy during interactions with the fake interviewer's voice.}
All 14 responded without hesitation to the three questions asked by
the fake interviewer, visited the requested website, and even gave 
their school ID number to the interviewers. After the interview concluded and the
deception was revealed, only four of the 14 participants 
  stated that they thought something was ``off'' about the fake
interviewer's voice. Importantly, these four participants had (intentionally)
not been given the ``priming'' statement that the fake interviewer
``had a cold.'' Below, we explore the most interesting results from
this study and highlight several key limitations.

{\bf \em 1) Reaction to fake voice: } Several themes arose during the post-deception
interviews, as summarized below. 
\begin{packed_itemize}
\item {\em Complete surprise: } Four participants 
    were visibly and audibly astonished when the deception was
  revealed. {\bf P5} noted that, ``I really thought it was you -- like
  100\%,'' while {\bf P10}, after a shocked moment of silence, said
   ``computers just won the Turing test.''
\item {\em Satisfied with ``sick'' excuse: } Seven participants 
    noted explicitly that the ``sick''
  excuse squashed any concerns about the fake interviewer's
  voice. {\bf P4} said that ``I think it totally worked -- I thought
  you were terribly sick,'' while {\bf P2} noted that ``it was really
  kind of worrying [how sick you sounded].''
\item \jenna{{\em Silently suspicious: } Four participants 
  ({\bf P9, P12, P13, P14}) 
  expressed suspicions {\em after} the deception was revealed.
  {\bf P12} and {\bf P13} said it ``sounded like {\em speaker} had a cold,'' and
  {\bf P14} supposed ``it was a poor quality microphone.''}
\end{packed_itemize}

{\bf \em 2) Why participants didn't voice their concerns: } After the
deception was revealed, participants were asked to
identify elements of the interview structure that increased their
trust in the fake interviewer. Some, of course, were completely
unsuspecting and did not think to question the fake
interviewer. However, others noted that 
the presence of a second (obviously human) interviewer, social
convention, and the origin of
the interview request \jenna{(from within our department)} bolstered their trust. 

\begin{packed_itemize}
\item {\em Presence of real interviewer: } Several
  participants credited the ``tag-team'' nature of the interview,
  with real and fake interviewers colluding, as making the deception more
  believable. ``I feel like [the real interviewer's] obviously human
  presence played a big factor in [my not saying anything].'' ({\bf
    P9}). 
\item {\em Polite social convention: } 
  Multiple participants noted that they felt it would be uncomfortable or wrong
  for them to say something about the fake interviewer's voice during
  the interviewer.
  When asked why they didn't say anything about the quality of the fake 
  interviewer's voice, {\bf P12} exclaimed, ``well that would be quite the insult!''  
\item {\em Provenance of interview request: } \jenna{Since
  we recruited from within our department, the recruitment was
  sent out through trusted channels only accessible by members of the 
  department (\ie email list-serv, Slack). {\bf P9} expressed suspicions
  during the debriefing, but credited the ``provenance of the study... seemed
  like a legit source'' as a reason to fully participate with our questions.}
\end{packed_itemize}

{\bf \em 3) What would have made participants suspicious: } When asked to
articulate what would have made them more suspicious, participant
responses varied.
\begin{packed_itemize}\vspace{-0.05in}
\item {\em Nothing: } Participants most surprised by the
  deception claimed that nothing would have caused them to question the
  credibility of the fake interviewer: ``I'm glad you guys didn't ask
  me for a bank account, because [...] I would have given it to
  you'' ({\bf P5}).  
\item {\em Requesting more personal information: } One participant noted ``I don't
  think the information you wanted was very sensitive [so] I don't see
  why I need to be concerned about this'' ({\bf P6}). IRB constraints
  prevented us from soliciting anything more personal than a  
  student ID,  used to access services at our  
  university. While not public, this information is not inherently
  sensitive. \vspace{-0.0in}
\end{packed_itemize}

\jenna{{\bf \em 4) Effect of familiarity with interviewers: } 
Seven of the participants 
rated their familiarity with both interviewers' voices as a 1 out of 5 
(\eg not at all familiar with either). 
Their responses, though, were consistent with the other participants who had 
some previous familiarity with one or both of the interviewers' voices. 
Only one participant ({\bf P8}) mentioned that ``the voice did seem pretty weird, 
but since I trust you both, I just went [with] it.'' 
These results suggest that the trusted setting and presence of a human likely 
play a larger factor than prior familiarity with the speaker's voice.}

{\bf \em 5) Effect of priming statement: } To examine the effect of the
``sick'' excuse on the believability of the fake voice, we conduct
four interviews in which participants are not told that the fake
interviewer is sick. In these interviews, participants exhibit an increased
level of skepticism about the fake interviewer during the debriefing. 
One claimed ``it was very obviously a fake voice,'' ({\bf P11}) but said that based on
their experience in other deception studies they decided not to say
anything. Others did not see through the deception but did
note that ``I was feeling weird'' ({\bf P13}) and ``I just feel your
voice is very strange'' ({\bf P14}).




\begin{table*}[t]
\centering
\resizebox{\linewidth}{!}{%
\begin{tabular}{c|c|l|l} 
\hline
\textbf{Category}~ & \textbf{Defense} & \multicolumn{1}{c|}{\textbf{Method}} & \multicolumn{1}{c}{\textbf{Limitations}}\\ 
\hline
\multirow{2}{*}{\begin{tabular}[c]{@{}c@{}}\textit{Liveness}\\\textit{Detection}\end{tabular}}
                   & \cite{zhang2017hearing} & Measures human vocal
                                               tract movement using
                                               Doppler radar.~ &
                                                                 Requires
                                                                 precise
                                                                 static
                                                                 calibration
                                                                 during
                                                                 enrollment/testing.
                                                                  \\ 
\cline{2-4}
 & \cite{wang2019voicepop} & ~Detects presence of human breath on mic.~ & Speaker must be  <4 inches from mic. \\ 
\hline
\multirow{2}{*}{\begin{tabular}[c]{@{}c@{}}\textit{Loudspeaker}\\\textit{Detection}\end{tabular}}
                   & \cite{chen2017you} & ~Detects presence of
                                          magnetic fields produced by
                                          loudspeakers. & Requires careful
                                                          motion of smartphone
                                                          during recording. \\ 
\cline{2-4}
 & \cite{catcher} & Compares audio environment to previously enrolled speaker environment. &                                                                  Requires
                                                                 precise
                                                                 static
                                                                 calibration
                                                                 during
                                                                 enrollment/testing. \\ 
\hline
\multirow{2}{*}{\begin{tabular}[c]{@{}c@{}}\textit{Artifact}\\\textit{Detection}\end{tabular}}
                   & \cite{gao2010audio, lieto2019hello, albadawy2019detecting, wang2020deepsonar,
                     ahmed2020void,lavrentyeva2017audio} & Trains models to recognize
                                      spectral characteristics of
                                      synthetic speech.~ & Only
                                                           effective
                                                           when audio
                                                           directly
                                                           played from
                                                           speaker. \\
\cline{2-4}
 & \cite{shirvanian2020voicefox} & Measures speech-to-text error rate of speech samples against ground truth. & Requires knowledge of ground truth audio content. \\ 
\hline
\begin{tabular}[c]{@{}c@{}}\textit{Preventing}\\\textit{Synthesis}\end{tabular}
                   & \cite{huang2020defending} & Corrupts speech
                                                 samples to prevent
                                                 unauthorized speech
                                                 synthesis. & Degrades
                                                              quality
                                                              of
                                                              defended
                                                              speech. \\
\hline
\end{tabular}
}
\caption{Taxonomy of defenses proposed to prevent speech synthesis
  attacks.}\vspace{-0.25in}
\label{tab:defenses}
\end{table*}

\vspace{-0.1in}
\subsection{Key Takeaways}

Our two user studies (A \& B) show that
context and demographics impact the credability of synthesized speech for \htrev{human users}. In study A, we found that mentioning
fake speech increased participants' skepticism of the fake speakers
they heard. Additionally, women and younger participants in study A
were more likely to correctly identify fake speakers.

Our key takeaway from study B is
that {\em a fake voice fooled humans in a trusted interview setting.}
Of particular interest is that all our study B participants were graduate
students in computer science, some of whom actively research 
security or machine learning. Our starting hypothesis 
was that computer science graduate students would be among the hardest targets
to fool with a fake voice. Yet, none of them expressed suspicion about
the fake voice during the interview.   

\para{Limitations \& Next Steps.} Our participant pool for study B was largely homogenous
in gender, age, and educational background.  \htrev{
  To conduct a ``trusted'' interview, our participants were drawn} 
from our academic department (computer science). The gender breakdown
of our participants matches that of the department, which skews
heavily male. \htrev{It is possible that the observed effect of gender and age on
responses in study A could also extend to study B.  Therefore, a
viable follow-up work is to conduct larger, more-diverse user studies to provide a more nuanced
understanding of synthesized voice attacks in trusted settings.



On a related note, our trusted interview in study B followed a voice-only 
  format, where voice is the only medium for interaction.  Yet in
  real-world scenarios, interviewees could use two-factor authentication
  mechanisms to verify the trusted setting, e.g.,  requesting the interviewers to turn
  on their video feed, or challenging the interviewers with some verbal
  tests. These combined verification methods could make the attacks much
  more difficult, allowing human users to effectively defend against 
  speech synthesis attacks.  We believe this is an important direction
  for follow-up work.


}

\vspace{-0.05in}
\section{Evaluating Existing Defenses}
\label{sec:existingdef}

Given the potency of these attacks, we now ask: ``what can be done to stop
them?'' Numerous defenses have been proposed to mitigate voice-based spoofing
attacks, many with significant assumptions that limit their practical applicability.
Here, we consider a range of defenses in light of our threat model and note
the limitations associated with different approaches.
Finally, we experimentally evaluate two representative defenses: one
that detects synthetic speech using physical artifacts from replay~\cite{ahmed2020void}, and
one that prevents voice synthesis by embedding audio
perturbations~\cite{huang2020defending}. 

\vspace{-0.05in}
\subsection{Existing Defenses: Detection \& Prevention}

We categorize existing defenses into multiple categories in
Table~\ref{tab:defenses}, noting approach and key limitations for each.

\para{Detecting Synthetic Speech/Speakers.} Most defenses counteract replay
or synthesis attacks by detecting certain artifacts, either environmental or
speech-specific. We further classify these defenses into three categories:
{\em liveness detection}, {\em loudspeaker detection}, or {\em synthesis
  artifact detection}.

The majority of detection defenses in Table~\ref{tab:defenses} make strong
assumptions about usage context. Some require a specific recording device,
others require precise placement of mic relative to speaker, or even careful
motion of the mic during recording, \emily{which raise significant
  accessibility and usability concerns}. These assumptions clearly limit the
applicability of these defenses in our real-world scenarios, such as those on
Alexa and WeChat. In most of these scenarios, defenders cannot control the position and movements of recording devices in the wild. 

\para{Preventing Unauthorized Voice Synthesis.} An orthogonal line of work
tries to prevent voice synthesis by embedding perturbations in audio samples
to shift them in the feature space~\cite{huang2020defending}.  It is the only
generalized defense to preventing synthesis attacks (rather than detecting
them post facto).


\para{Defenses We Evaluate.} From the defenses listed in
Table~\ref{tab:defenses}, we choose to evaluate two representative
defenses. From the set of detection defenses, we choose
Void~\cite{ahmed2020void} due to its recency, high performance, and
relatively few operating constraints. Void uses low spectral frequencies to
distinguish human speech from speech replayed through loudspeakers.  We also
evaluate Attack-VC~\cite{huang2020defending}, the only system we're aware of
that prevents voice synthesis. \rev{Finally, we also evaluate when
  Void and Attack-VC are combined together to make a stronger defense. }



\subsection{Detecting Synthetic Speech using Void}

Void~\cite{ahmed2020void} protects {\em systems} against
synthesized speech attacks. It identifies $97$ distinct low-frequency
spectral features that distinguish human speech and replayed speech. These features can be used to train a variety of detection
models. Since the WeChat/Alexa attacks rely on replayed synthesized
speech, this defense applies to our setting. In the original paper, Void is
tested extensively on replay attacks but only cursorily on synthesis attacks. 

\para{Methodology.} We recreate the feature extraction pipeline
of~\cite{ahmed2020void} and train three models using
the 2017 ASVSpoof dataset~\cite{kinnunen2017asvspoof}. Like Void, we
report the equal error rate ($EER$) for each of the trained models,
and the detection success rate (i.e. model's ability to
distinguish real/replayed speech). $EER$ measures when the false positive and false negative rates of a system are
equal and is commonly used to report performance of biometric
systems. The $EER$ of our trained models is on par with that reported in
the original paper.

\begin{packed_itemize}
  \item {\bf Support Vector Machine (SVM) with RBF Kernel}: This model
    had the best reported performance in~\cite{ahmed2020void}. 
  \item {\bf LightCNN}: \cite{lavrentyeva2017audio} propose a
    27-layer DNN for synthetic speech detection which is also evaluated in~\cite{ahmed2020void}. We
    use the same architecture and parameters as
    in~\cite{lavrentyeva2017audio}, but modify the input size to
    accommodate~\cite{ahmed2020void}'s 97 features. 
  \item {\bf Custom CNN:} Our last model is a custom 5-layer
    CNN (see Table~\ref{tab:custom_dnn} in the Appendix). We train
    this model for 25 epochs using the Adam optimizer with
    $lr=0.001$. 
  \end{packed_itemize}
  
We test the models on a custom dataset of
replayed synthesized samples (targeting VCTK speakers).
The synthesized samples are generated on SV2TTS with
$N=20$ source files per speaker. They are replayed over two different
devices (a UE Boom loudspeaker and LG UltraFine 4K
monitor) and recorded using an iPhone 11 situated 6 inches from the audio
source. Each replayed set contains $200$ samples from $20$ different
speakers. For comparison, we add $200$ clean samples from the same
speakers to each replayed set.

\para{Results.} Void reliably differentiates real and synthesized samples in
our two custom datasets but has high EERs (i.e. false
positive/negative rate) across all models, as shown in Table~\ref{tab:void_results}. All
models have $>88\%$ detection success rate, but $EER$ for all models
is $>5\%$. High-performing biometric systems typically have $EER<1\%$.

\begin{table}[t]
\centering
\resizebox{\linewidth}{!}{%
\begin{tabular}{l|c|c|c||c|c|c} 
\hline
\multirow{2}{*}{\diagbox{\begin{tabular}[c]{@{}l@{}}\textbf{Replay}\\\textbf{Source}\end{tabular}}{\textbf{Metric}}} & \multicolumn{3}{c||}{\textit{Detection Success~}} & \multicolumn{3}{c}{\textit{EER}} \\ 
\cline{2-7}
 & \textbf{SVM} & \textbf{LightCNN} & \begin{tabular}[c]{@{}c@{}}\textbf{Custom }\\\textbf{DNN}\end{tabular} & \textbf{SVM} & \begin{tabular}[c]{@{}c@{}}\textbf{Light}\\\textbf{CNN}\end{tabular} & \begin{tabular}[c]{@{}c@{}}\textbf{Custom}\\\textbf{DNN}\end{tabular} \\ 
\hline
\multicolumn{1}{c|}{\textbf{UE Boom}} & 88.5\% & 90.9\% & 92.0\% & 12.1\% & 9.0\% & 7.5\% \\ 
\hline
\multicolumn{1}{c|}{\textbf{LG Monitor}} & 89.3\% & 89.3\% & 91.8\% & 5.1\% & 9.5\% & 7.5\% \\
\hline
\end{tabular}
}
\caption{Results from running 3 Void detection models on our two synthetic
speech datasets.}
\label{tab:void_results}
\vspace{-0.3in}
\end{table}

\para{Discussion.} Void's high $EER$ renders it less effective in
practice in our setting, although the original paper reports a much
lower $EER$ when using a custom training dataset. If the custom training dataset were
more widely available, Void could provide effective protection for
scenarios like the WeChat/Alexa attacks (\S\ref{sec:blackbox}).


\subsection{Preventing Speech Synthesis via Attack-VC}

Attack-VC~\cite{huang2020defending} is designed to protect users from having
their voice copied via speech synthesis. Attack-VC
adds carefully designed perturbations to speech samples that disrupt
unauthorized future synthesis. The ``embedding'' perturbation generation method
in~\cite{huang2020defending} assumes full knowledge of the
downstream voice synthesis model $\mathcal{M}$ (\ie a white-box threat
model). A defender uses the speaker embedding component of 
$\mathcal{M}$ to create a size-bounded perturbation $\delta$ that shifts the speaker embedding of their sample $x$
towards the embedding of a different speaker's sample $d$. Then, an
adversary \adv{} who steals the victim \targ{}'s defended sample $x + \delta$, cannot use $\mathcal{M}$ to successfully synthesize a
fake voice sample. The synthesized samples \sadv{} should not sound like \targ{}. 

\para{Methodology.} We perform a small-scale study using the VCTK dataset and two models --
AutoVC and SV2TTS (as in~\ref{sec:whitebox}). We use the same subset
of 20 VCTK speakers as in \S\ref{sec:whitebox}. Using author-provided code~\cite{github_attackvc},
we generate 19 defended samples per speaker (using the other speakers
as optimization targets). 
We test three perturbation levels, $\epsilon = 0.01, 0.05, 0.1$,
following~\cite{huang2020defending}.

\begin{figure}[t]
  \centering
  \includegraphics[width=0.45\textwidth]{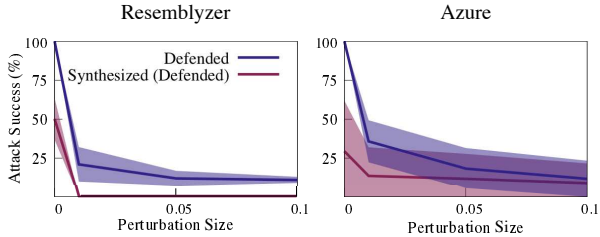}
  \vspace{-0.15in}
  \caption{Performance of both defended and ``synthesized from
    defended'' samples on Resembylzer and Attack-VC. These samples
    were protected and synthesized using SV2TTS. Results on AutoVC are
    similar.}
  \label{fig:attack_vc}
\end{figure}

We notice that the original perturbation loss
function $\mathcal{L}$ from~\cite{huang2020defending} does not sufficiently constrain
perturbation size. This results in large perturbations that make
defended audio samples sound inhuman. To fix this,
we add a term to $\mathcal{L}$ (new term is {\bf bold}):
\begin{equation}
\mathcal{L} = \alpha \cdot MSE(x + \delta, d) - \beta \cdot MSE(x +
\delta, x) + \boldsymbol{\gamma \cdot \left\lVert \delta \right\rVert},
\end{equation}
where $MSE$ represents the mean-squared error. This additional term
makes the perturbation less audible but does not
affect attack success. Empirically, setting $\alpha,
\beta=1$ and $\gamma=0.1$ works best. We multiply $\gamma$ by $0.99$
every $100$ iterations.

Then, we use the methodology of \S\ref{sec:whitebox} to synthesize
speech from the defended samples. Both the defended
samples and the ``synthesized-from-defended'' samples are evaluated
against Azure and Resemblyzer (see \S\ref{sec:machines} for details).

\para{Results.} As Figure~\ref{fig:attack_vc} shows, Attack-VC
does thwart voice synthesis, but it also corrupts defended samples
beyond reliable recognition. For both models and all speaker recognition
systems, the speaker recognition accuracy
for``synthesized-from-defended'' samples is less than $35\%$, meaning
that synthesis attacks after Attack-VC are less successful. However,
speaker recognition accuracy for ``defended'' samples is at most
$55\%$, meaning that they cannot be properly matched to the true
speaker. Additionally, ``defended'' samples still have significant
audible distortion, even with our additional constraints on
perturbation size.

\rev{
  \subsection{Combining Void and Attack-VC}

 Finally, we evaluate a ``stronger'' defense that combines Void and
 Attack-VC, but find that it only provides marginal benefits.  \rev{In this
   experiment},  we test Void's detection
efficacy on speech synthesized from Attack-VC protected samples, with  varying perturbation levels $\epsilon=0.01, 0.05, 0.1$.  We find that speech generated from protected samples
with $\epsilon < 0.1$ can only be detected $2-4\%$ better (with lower EERs) than
normal synthetic speech. 
Detailed results are in the Appendix. }

\rev{

\subsection{Key Takeaways}



\htrev{Our results} demonstrate a significant need for new and improved defenses against synthesized
speech attacks, particularly defenses generalizable enough for
real-world applications. While Void reliably detects fake speech played through speakers, its applicability is limited to replay
attacks. Meanwhile, existing prevention defenses such as Attack-VC distort voices beyond recognition, and might benefit from using acoustic hiding techniques~\cite{qin2019imperceptible}. \htrev{These defenses} also assume perfect
(white-box) knowledge of the attacker's speech synthesis model, which is unrealistic in
real-world settings.

\para{Limitations \& Next Steps.} We only evaluate two
representative and \htrev{top-performing} defenses (one for each category) and their combined
effect. A more comprehensive investigation is required, especially as
new defenses emerge.


\htrev{We also note that current defenses focus on protecting  SR systems.} However,
our results in \S\ref{sec:trusted} indicate an equal need for
human-centric defenses against synthetic speech. \htrev{One possible
  direction is to make synthetic
  speech more ``obvious'' to human audiences,} either by corrupting
its generation process to make the speech sound inhuman
(i.e., Attack-VC's yet-unreached goal) or \htrev{designing} parallel authentication methods
(i.e. video feed or vocal challenges) that help expose fake speakers.  
}

\rev{
\section{Conclusion}
\label{sec:discussion}

Our work represents a first step towards understanding
the real-world threat of deep learning-based speech synthesis attacks.  Our
results demonstrate that synthetic speech generated using publicly available 
systems can already fool both humans and today's popular software systems,
and that existing defenses fall short.  As such, our work highlights the need for new
defenses, for both humans and machines, against speech synthesis
attacks, promote further research efforts for exploring subsequent challenges and
opportunities, while providing a solid benchmark for future research. 



}

\section*{Acknowledgements}
We thank our anonymous reviewers for their insightful feedback. This work is supported in part by NSF grants CNS-1949650, CNS-1923778,
CNS1705042, and by the DARPA GARD program. Emily Wenger is also supported by a GFSD fellowship. Any
opinions, findings, and conclusions or recommendations expressed in this material are those of the authors and do not
necessarily reflect the views of any funding agencies.

\bibliographystyle{ACM-Reference-Format}
\bibliography{references}

\newpage
\section{Appendix}

\subsection{Methodology for \S 4.1}

Here, we describe the methodology used to
recreate~\cite{mukhopadhyay2015all} in \S 4.1.

\para{Systems Used.} We use the following synthesis and SR systems.

\begin{packed_itemize}
\item {\bf Festvox} ~\cite{festvox} is a text-dependent voice conversion
   system that uses GMMs \cite{reynolds2000speaker} to model the vocal characteristics.
   Training the system requires 4-8 minutes of speech from both the
   source and target speakers and takes approximately 5 - 10 minutes.
 \item {\bf Bob Spear} ~\cite{khoury2014spear} is an open-source speaker
   recognition system. It uses classical statistical techniques
   (Universal Background Model Gaussian Mixture Models (UBM-GMM) and
   Inter-Session Variability (ISV)) to perform speaker
   recognition~\cite{vogt2008explicit}.
 \item {\bf Resemblyzer}, {\bf Azure} See \S\ref{sec:threat}.
 \end{packed_itemize}
 
\para{Datasets Used.} The prior attack uses the CMU ARCTIC and Voxforge
datasets. CMU ARCTIC~\cite{arctic} contains
1150 spoken phrases from male and female English speakers. 
Voxforge~\cite{voxforge} is an open-source dataset of transcribed
speech. The subset we use contains 6561 spoken phrases, each
approximately 5 seconds long, from 30 English speakers.

 \para{Attack Implementation.} We recreate the exact setup of the voice conversion attack from
~\cite{mukhopadhyay2015all}. We use
Festvox to convert speech from two CMU ARCTIC speakers (one male, one female) to imitate 10 speakers of the Voxforge dataset. 
The synthesized attack samples are tested against both of Bob Spear's
speaker recognition algorithms (UBM-GMM and ISV), along with
Resemblyzer and Azure. We measure attack performance using the {\em attack
  success rate (AS)}, which denotes the percent of synthesized
samples identified as the target.
  
\para{Attack Results.}  Our attack results on Bob Spear align with
 prior work, with an average $AS$ of
$93.1\%$ for Spear's UBM-GMM method and $97.1\%$ for Spear's ISV method (see
Table~\ref{tab:festvox}). Our synthesized samples have an average
Mel-Cepstral Distortion (MCD) of 4.79 dB, compared to the average MCD of 5.59 dB
reported in~\cite{mukhopadhyay2015all} (lower is better).

However, the Festvox attack {\em fails on current SR systems}, including Resemblyzer
and Azure. Out of the $38$ speakers tested, only $8$ have one or more
attack samples accepted ($21\%$). Moreover, Festvox requires a significant amount (around 6-8 minutes) of content-specific speech
from both the victim and attacker to synthesize fake speech, making
the attack impractical. Modern voice conversion systems like SV2TTS are text-independent
and need much fewer victim speech samples to successfully synthesize
speech. 

\subsection{The Rainbow Passage}
\label{sec:rainbow}

Participants in our user study of \S 4.2 were asked to
record the Rainbow Passage. The Rainbow Passage is commonly used in
linguistic studies, since it contains nearly all the phonemes
combinations of the English language. The full text of the
Rainbow Passage is below:

\begin{displayquote}
When sunlight strikes raindrops in the air, they act like a prism and
form a rainbow. The rainbow is a division of white light into many
beautiful colors. These take the shape of a long round arch, with its
path high above, and its two ends apparently beyond the horizon. There
is, according to legend, a boiling pot of gold at one end. People look
but no one ever finds it. When a man looks for something beyond his
reach, his friends say he is looking for the pot of gold at the end of
the rainbow. Throughout the centuries men have explained the rainbow
in various ways. Some have accepted it as a miracle without physical
explanation. To the Hebrews it was a token that there would be no more
universal floods. The Greeks used to imagine that it was a sign from
the gods to foretell war or heavy rain. The Norsemen considered the
rainbow as a bridge over which the gods passed from the earth to their
home in the sky. Other men have tried to explain the phenomenon
physically. Aristotle thought that the rainbow was caused by
reflection of the sun's rays by the rain. Since then physicists have
found that it is not the reflection, but refraction by the raindrops
which causes the rainbow. Many complicated ideas about the rainbow
have been formed. The difference in the rainbow depends considerably
upon the size of the water drops, and the width of the colored band
increases as the size of the drops increases. The actual primary rainbow observed is said to be the effect of
superposition of a number of bows. If the red of the second bow
falls upon the green of the first, the result is to give a bow with an
abnormally wide yellow band, since red and green light when
mixed form yellow. This is a very common type of bow, one
showing mainly red and yellow, with little or no green or blue.
\end{displayquote}

\subsection{Phrases for Synthesis}
\label{sec:phrases}

Table~\ref{tab:phrases} lists the phrases used for synthetic speech
generation with SV2TTS and AutoVC.

\begin{table}[h]
\centering
\resizebox{0.5\textwidth}{!}{%
\begin{tabular}{c}
  \textbf{Phrases used for SV2TTS Speech Synthesis} \\ 
  \hline
  \begin{tabular}[c]{@{}c@{}}We control complexity by establishing new languages for describing a design, \\each of which emphasizes particular aspects of the design and deemphasizes others.\end{tabular} \\ 
  \hline
  An interpreter raises the machine to the level of the user program. \\ 
  \hline
  Everything should be made as simple as possible, and no simpler.\\
  \hline
  \begin{tabular}[c]{@{}c@{}}The great dividing line between success and failure \\can be expressed in five words: "I did not have time."\end{tabular} \\ 
  \hline
  \begin{tabular}[c]{@{}c@{}}When your enemy is making a very serious mistake, \\don't be impolite and disturb him.\end{tabular} \\ 
  \hline
  \begin{tabular}[c]{@{}c@{}}A charlatan makes obscure what is clear; \\a thinker makes clear what is obscure.\end{tabular} \\ 
\hline
  \begin{tabular}[c]{@{}c@{}}There are two ways of constructing a software design; \\one way is to make it so simple that there are obviously no deficiencies,\\and the other way is to make it so complicated that there are no obvious deficiencies.\end{tabular} \\ 
  \hline
  The three chief virtues of a programmer are: Laziness, Impatience and Hubris. \\ 
  \hline
All non-trivial abstractions, to some degree, are leaky. \\ 
  \hline
  XML wasn't designed to be edited by humans on a regular basis. \\
  \hline
\end{tabular}
}
\caption{Phrases used for SV2TTS speech synthesis attacks in \S 4.2 and 4.3.}
\label{tab:phrases}
\end{table}

\subsection{Phoneme Similarity Experiments from \S 4.2}

\begin{table}
\centering
\resizebox{0.49\textwidth}{!}{%
\begin{tabular}{c|c} 
\hline
PER  & Text \\ 
\hline
0.0  & \begin{tabular}[c]{@{}c@{}}wind had come up and was driving puffy white clouds across the sky\end{tabular}                                \\
0.25 & \begin{tabular}[c]{@{}c@{}}wind had shove pup tuned wess priming puffy weir cloudy across the ski\end{tabular}                            \\
0.5  & \begin{tabular}[c]{@{}c@{}}wind had kerl yum econocom gestate erring puffy lett clouds close aja sky\end{tabular}                         \\
0.75 & \begin{tabular}[c]{@{}c@{}}vine hyde tonne edmundson airships stennis orville gaffes sues lader kross the sky\end{tabular}                \\
1.0  & \begin{tabular}[c]{@{}c@{}}vine tass klumb muhs caftans prognosticater mendiola morphogenesis clips hilder kross though ski\end{tabular}  \\
\hline
\end{tabular}%
}
\caption{Changes to a reference text (top line) as phoneme error rate (PER)
increases. Changes are generated by randomly selecting phrases from
CMUDict~\cite{cmu_dict} which match the specified PER.}
\label{tab:per}
\end{table}

We use phoneme error rate (PER) to determine the
similarity between a target sample's content and a synthesized
output's content. PER is the Levenshtein 
distance between a predicted and reference pronunciation's phonemes
and is a common measure of phonetic similarity between two text
samples~\cite{bisani2008g2p}.

To control the phonetic distance between sample and output, we implement
a custom phoneme conversion system, which takes in a transcript of the
target speech sample and produces output text with a specific phoneme
error rate (PER) from the input.  Output text was generated for 
PERs normalized by phoneme length ranging from $0.0$ to $1.0$. Using SV2TTS, we synthesize
speech from the output text.

Table~\ref{tab:per} demonstrates how different phoneme error rates affects the 
output text. To achieve a particular phoneme error rate, a random distribution 
of PER per word, given the total PER for the reference text, was generated. Words
satisfying this distance criteria from the original text were then randomly 
selected from CMUDict to generate the output text ~\cite{cmu_dict}. Using
1 speech sample, we find that the PER between the target sample content and 
synthesized sample content does not impact attack success 
(see Table~\ref{tab:phonemeresults}).


\begin{table}
\centering
\resizebox{0.49\textwidth}{!}{%
\begin{tabular}{c|l|lllll|lllll} 
\hline
\multicolumn{1}{c}{}                  &          & \multicolumn{5}{c|}{\textbf{LibriSpeech}}                                                      & \multicolumn{5}{c}{\textbf{VCTK}}                                                                                 \\ 
\cline{3-12}
\multicolumn{1}{l}{}                  &          & \multicolumn{1}{c}{\textit{0.0}} & \textit{0.25} & \textit{0.5} & \textit{0.75} & \textit{1.0} & \multicolumn{1}{c}{\textit{0.0}}                   & \textit{0.25} & \textit{0.5} & \textit{0.75} & \textit{1.0}  \\ 
\hline
\multirow{2}{*}{\textbf{Resemblyzer}} & $\mu$    & $100\%$                          & $100\%$       & $100\%$      & $100\%$       & $100\%$      & \begin{tabular}[c]{@{}l@{}}$22.2\%$\\\end{tabular} & $15.5\%$      & $13.0\%$     & $15.0\%$      & $20.0\%$      \\
                                      & $\sigma$ & $0\%$                            & $0\%$         & $0\%$        & $0\%$         & $0\%$        & $6.4\%$                                            & $6.8\%$       & $4\%$        & $5.5\%$       & $8.1\%$       \\ 
\hline
\multicolumn{2}{c|}{\textbf{Azure}}              & $18\%$                           & $19.5\%$      & $16\%$       & $19\%$        & $13.5\%$     & $12\%$                                             & $15\%$        & $15.5\%$     & $15.5\%$      & $18.5\%$      \\
\hline
\end{tabular}%
}
\caption{Resemblyzer, Azure results for phoneme experiments. Results on Resemblyzer reported as average, standard deviation of $10$ trials.}
\label{tab:phonemeresults}
\end{table}

\rev{
  \subsection{Post-Deception Interview Procedure from \S 5.3}

  Here, we describe the script and format of our post-deception
  interview from \S 5.3.

  \para{Script.} The following questions were asked during the
  post-deception interview:

  \begin{packed_itemize}
    \item How familiar were you with <fake interviewer's> voice before
      this study (on a scale of 1-5, 5 being most familiar)?
    \item How familiar were you with <real interviewer's> voice before
      this study (on a scale of 1-5, 5 being most familiar)?
    \item Did anything throughout the interview cause you to think
      that <fake interviewer> wasn't using a real voice?
    \item Were you at any point suspicious of the interview setup?
    \item If you were, at what point did you become suspicious?
    \item What, if anything, would have made you suspicious of <fake
        interviewer>?
    \item Do you have any additional comments about your
          experience?      
  \end{packed_itemize}

  \para{Qualitative Analysis Procedure.}
Following best practices for analysis of open-ended questions, as described in ~\cite{codingmanual},
the interviewers made notes during and/or immediately following the post-deception portion 
of the interview. After the conclusion of all interviews, each interviewer independently 
reviewed their notes and/or rewatched the interviews as needed. 
For some questions, \ie level of familiarity with interviewers, the responses already had discrete categories.
For open-ended questions, using a bottom-up approach, the notes were used to categorize responses into
general themes regarding the interviewees level of suspicion, and reasons why. 
Upon completing independent coding of all interviews, the researchers met to consolidate
the code books into a single code book with consistent categories, and resolve any discrepancies.
In the end, each coding category represents sentiments expressed by 3 or more participants,
unless otherwise noted in \S 5.3.


}

\rev{

  \subsection{Results on Combined Defenses from \S 6}

We also tested if Void can detect speech synthesized
from AttackVC-protected samples. We first generated protected samples using $\epsilon=0.01, 0.05, 0.1$ for
AttackVC (as in \S 6.1), then recorded them using the UE-Boom
loudspeaker for playback as before. When tested, all three Void models
(SVM/LightCNN/CustomDNN) show high detection rates (91-94\%) on
protected samples at $\epsilon=0.01, 0.05$ levels but slightly lower
detection rates at $\epsilon=0.1$. Detection rates for protected samples
with $\epsilon < 0.1$ are higher than those for unprotected samples (see
Table 9).
}

\begin{table}
\centering
\resizebox{\linewidth}{!}{%
  \rev{
\begin{tabular}{l|c|c|c||c|c|c} 
\hline
\multirow{2}{*}{\diagbox{\begin{tabular}[c]{@{}l@{}}\textbf{Replay}\\\textbf{Source}\end{tabular}}{\textbf{Metric}}} & \multicolumn{3}{c||}{\textit{Detection Success~}} & \multicolumn{3}{c}{\textit{EER}} \\ 
\cline{2-7}
 & \textbf{SVM} & \textbf{LightCNN} & \begin{tabular}[c]{@{}c@{}}\textbf{Custom }\\\textbf{DNN}\end{tabular} & \textbf{SVM} & \begin{tabular}[c]{@{}c@{}}\textbf{Light}\\\textbf{CNN}\end{tabular} & \begin{tabular}[c]{@{}c@{}}\textbf{Custom}\\\textbf{DNN}\end{tabular} \\ 
\hline
\multicolumn{1}{c|}{\textbf{$\epsilon = 0.01$}} & 91.3\% & 94.1\% & 90.2\% & 4.8\% & 3.1\% & 9.8\% \\ 
\hline
\multicolumn{1}{c|}{\textbf{$\epsilon = 0.05$}} & 93.0\% & 94.6\% & 91.4\% & 3.9\% & 4.4\% & 8.3\% \\
  \hline
\multicolumn{1}{c|}{\textbf{$\epsilon = 0.10$}} & 90.8\% & 80.1\% & 84.4\% & 6.2\% & 19.1\% & 17.9\% \\
\hline
\end{tabular}
}
}
\caption{\rev{Void detection success rates and EER for combined
  defenses. Speech is first synthesized from Attack-VC defended
  samples (different $\epsilon$ levels indicated by row), replayed using
  the UE Boom loudspeaker, then run through the Void models.}}
\label{tab:void_results_combined}
\vspace{-0.25in}
\end{table}

\subsection{DNN Architecture from \S 6.2}

Table~\ref{tab:custom_dnn} lists the architecture used for our
custom DNN trained for Void in \S 6.2.

\begin{table}[h]
  \centering
  \resizebox{0.49\textwidth}{!}{%
    \begin{tabular}{crrccrr} 
      \hline
      \begin{tabular}[c]{@{}c@{}}\textbf{Layer }\\\textbf{Index}\end{tabular} & \multicolumn{1}{c}{\begin{tabular}[c]{@{}c@{}}\textbf{Layer}\\\textbf{Name}\end{tabular}} & \multicolumn{1}{c}{\begin{tabular}[c]{@{}c@{}}\textbf{~Layer }\\\textbf{~Type~}\end{tabular}} & \begin{tabular}[c]{@{}c@{}}\textbf{\# of }\\\textbf{Channels}\end{tabular} & \begin{tabular}[c]{@{}c@{}}\textbf{Filter}\\\textbf{Size~}\end{tabular} & \multicolumn{1}{c}{\textbf{Activation}} & \multicolumn{1}{c}{\begin{tabular}[c]{@{}c@{}}\textbf{~Connected }\\\textbf{to}\end{tabular}} \\ 
\hline
      1 & conv\_1 & Conv1D & 32 & 3 & ReLU &  \\
      2 & conv\_2 & Conv1D & 64 & 3 & ReLU & conv\_1 \\
      3 & max\_1 & MaxPool & 64 & 2 & - & conv\_2 \\
      4 & fc\_1 & FC & 128 & - & ReLU & max\_1 \\
      5 & fc\_2 & FC & 2 & - & Softmax & fc\_1 \\
      \hline
    \end{tabular}
  }
  \caption{Custom DNN architecture used to evaluate the Void defense in~\S 6.2.}
  \label{tab:custom_dnn}
\end{table}

\end{document}